\documentclass[11pt,article,nofootinbib,notitlepage]{revtex4-1}
\usepackage{amsmath,amssymb,float,color}
\usepackage[pdftex]{graphicx}

\def\bx{{\boldsymbol x}}

\def\bq{{\boldsymbol q}}
\def\bp{{\boldsymbol p}}
\def\bv{{\boldsymbol v}}

\def\xperp{x_{\!\perp}} 
\def\pperp{p_{\!\perp}}
\newcommand{\ltsim}{\protect\raisebox{-0.5ex}{$\:\stackrel{\textstyle <}{\sim}\:$}}
\newcommand{\gtsim}{\protect\raisebox{-0.5ex}{$\:\stackrel{\textstyle >}{\sim}\:$}}

\begin{document}

\title{Parametric estimate of the relative photon yields from the Glasma and the Quark-Gluon Plasma in heavy-ion collisions}
\author{J\"{u}rgen Berges}
\affiliation{Institut f\"{u}r Theoretische Physik, Universit\"{a}t Heidelberg, Philosophenweg 16, 69120 Heidelberg, Germany}
\author{Klaus Reygers}
\affiliation{Physikalisches Institut, Im Neuenheimer Feld 226, 69120 Heidelberg, Germany}
\author{Naoto Tanji}
\email{tanji@thphys.uni-heidelberg.de}
\affiliation{Institut f\"{u}r Theoretische Physik, Universit\"{a}t Heidelberg, Philosophenweg 12, 69120 Heidelberg, Germany}
\author{Raju Venugopalan}
\affiliation{Physics Department, Brookhaven National Laboratory, Bldg. 510A, Upton, NY 11973, USA}
\date{\today}

\begin{abstract}
Recent classical-statistical numerical simulations have established the ``bottom-up" thermalization scenario of Baier et al.~\cite{Baier:2000sb} as the correct weak coupling effective theory for thermalization in ultrarelativistic heavy-ion collisions. We perform a parametric study of photon production in the various stages of this bottom-up framework to ascertain the relative contribution of the off-equilibrium ``Glasma"  relative to that of a thermalized Quark-Gluon Plasma. Taking into account the constraints imposed by the measured charged hadron multiplicities at RHIC and the LHC, we  find that Glasma contributions are important especially for large values of the saturation scale at both energies. These non-equilibrium effects  should therefore be taken into account in studies where weak coupling methods are employed to compute photon yields. 
\end{abstract}

\maketitle

\section{Introduction} \label{sec:intro}

Significant theoretical progress in understanding the space-time evolution of ultrarelativistic heavy-ion collisions can be achieved in the idealized high-energy limit where the QCD coupling $\alpha_s \ll 1$. The ab initio dynamics of such a system shortly after the collision corresponds to that of an over-occupied, strongly correlated non-Abelian plasma, exploding into the vacuum along the beam axis of the colliding nuclei~\cite{Kovner:1995ja,Krasnitz:1998ns,Gelis:2007kn}. The properties of this over-occupied plasma of strongly correlated quarks and gluons, often called a Glasma~\cite{Lappi:2006fp}, can be determined by employing classical-statistical methods. 

An unfortunate complication is that the ab initio classical-statistical framework breaks down when the gluon occupancy becomes of order unity and the ``quantum one-half" contributions become comparable to the leading classical contributions in real-time correlation functions~\cite{Aarts:1997kp,Berges:2013lsa}. When this occurs, quantum kinetic descriptions are appropriate. However, because there is a significant overlap between the classical and quantum regimes in the evolution of the Glasma~\cite{Mueller:2002gd,Jeon:2004dh}, classical-statistical simulations can help identify the right effective kinetic theory for the subsequent evolution of the Glasma to thermal equilibrium.  The proper matching of the two frameworks is essential because the complexity of the dynamics of infrared modes in the system can lead to a number of weak coupling kinetic thermalization scenarios~\cite{Baier:2000sb,Blaizot:2011xf,Kurkela:2011ti,Bodeker:2005nv}. 

A recent breakthrough was achieved through large-scale numerical simulations of expanding non-Abelian plasmas in weak coupling where it was demonstrated that the Glasma flows to a non-thermal fixed point~\cite{Berges:2013eia,Berges:2013fga} that is insensitive to details of the initial conditions\footnote{It is worth noting that Color Glass Condensate (CGC)~\cite{Gelis:2010nm} initial conditions for the Glasma~\cite{Epelbaum:2013waa} lead very rapidly to a gluon number over-occupancy~\cite{Berges:2014yta} that subsequently flows to this non-thermal fixed point.}. Remarkably, the non-thermal fixed point identified by the classical-statistical simulations corresponds to the early stage of the ``bottom-up" thermalization scenario of Baier et al.~\cite{Baier:2000sb}--henceforth referred to by the acronym BMSS. This result was unanticipated because it was believed previously~\cite{Arnold:2003rq} that plasma instabilities (not included in the BMSS framework) should in principle play a big role in kinetic realizations of the expanding Glasma~\cite{Mrowczynski:1993qm,Rebhan:2005re,Romatschke:2005pm,Mrowczynski:2016etf}. The puzzling absence of late time plasma instabilities in the Glasma is strongly indicative of the large role of infrared modes as suggested by numerical results on the longitudinal to transverse pressure ratio~\cite{Berges:2015ixa} and in the striking universality of the  non-Abelian non-thermal fixed point to that exhibited by expanding self-interacting scalar fields~\cite{Berges:2014bba}. Thus the BMSS kinetic theory--in the regime where occupancies are large--is best regarded as an effective description that captures the correct physics of the Glasma, in analogy to effective kinetic descriptions of weak wave turbulence~\cite{Zakharov-book}.

Given the significant developments we outlined in the context of real-time studies of early times, it is important to understand their phenomenological consequences for heavy-ion collisions. These can be ascertained by extrapolating the weak coupling results to realistic computations. For non-thermal fixed points in scalar theories, it has been shown recently that such extrapolations of the classical-statistical results are robust even for values of the scalar coupling constant $\lambda\sim 1$~\cite{Berges:2016nru}. For gauge theories, the validity of such extrapolations is open to question; it is nevertheless useful to perform such an extrapolation and understand the phenomenological consequences thereof. Adopting this point of view, it has been shown recently~\cite{Kurkela:2015qoa,Keegan:2016cpi} that a sophisticated implementation of the effective kinetic theory~\cite{Arnold:2002zm} can be smoothly matched to relativistic viscous hydrodynamics on the early time scales required by heavy-ion phenomenology~\cite{Gale:2013da}. 

Electromagnetic signatures such as photon production are uniquely sensitive to the different stages of evolution in heavy-ion collisions. The relative rates of their production from the Quark-Gluon Plasma (QGP) and hadron gas stages of the evolution have been discussed for some time~\cite{Kapusta:1991qp,Shen:2013vja}. The measurement of the elliptic flow coefficient for photons (which arise of course from the underlying anistropic flow of quarks and hadrons~\cite{Chatterjee:2005de}) adds another handle to  probe the space-time dynamics at the time of emission~\cite{Adare:2015lcd,Adam:2015lda}. Comparisons of hydrodynamic and transport models, which implement contributions to the photon yields and elliptic flow from both thermal QGP emission and from the hadron gas, to the available data, indicate that reproducing both the photon yield and elliptic flow simultaneously is challenging in these models~\cite{Chatterjee:2013naa,Shen:2014lpa,vanHees:2014ida,Bratkovskaya:2014mva,Paquet:2015lta}. This discrepancy has been dubbed the ``direct photon puzzle"--for a recent review, see \cite{Shen:2016odt}. 

The above estimates do not include the contribution from pre-equilibrium photon production\footnote{For a discussion in strong coupling frameworks, see~\cite{Iatrakis:2016ugz}. For a recent discussion in the context of partonic transport simulations, see~\cite{Greif:2016jeb}.}. These may be especially  important in semi-peripheral heavy-ion collisions and in proton-nucleus collisions, where their relative contribution to the thermal yield~\cite{Shen:2015qba},  as a function of centrality, may help constrain the onset of thermalization in QCD matter. First phenomenological studies of photon production in the Glasma~\cite{McLerran:2014hza,McLerran:2014oea} suggest that this contribution is significant. In particular, it is argued~\cite{McLerran:2015mda} that photon yields and elliptic flow can be explained if the Glasma has a hard component which thermalizes at late times relative to thermalization times typically assumed in hydrodynamic simulations. 

In this work, we will estimate the rate of photon production within the BMSS framework. In particular, we will extract the parametric dependence of the rate, from different stages of the evolution of the Glasma, on the QCD coupling constant. These rates will be compared to that of photon production from the equilibrated QGP. We find that our conclusions are very sensitive to the saturation scale $Q_s$ in the Glasma and coefficients that relate this scale to the initial temperature and thermalization time. These coefficients can be fixed to fair accuracy from heavy-ion data on charged particle multiplicites; it is nevertheless subject to significant systematic uncertainties that we will elaborate on. 

As noted, the results quoted in this paper are parametric estimates and valid only in the kinetic regime. The results are however interesting enough to suggest that a more detailed computation of pre-equilibrium photon production rates is desirable. When gluon occupancies are large, this can be achieved by using classical-statistical simulations including dynamical quarks \cite{Gelis:2005pb,Gelis:2015eua,Gelfand:2016prm}\footnote{Similar computations have been performed in proton-nucleus collisions~\cite{Gelis:2002ki,Benic:2016yqt,Benic:2016uku}. In this case, at least for minimum-bias collisions, Glasma evolution is not significant.}. First results for the photon production from color background fields in a fixed box geometry are now available \cite{Tanji:2015ata}. When the occupancy of gluons becomes of order unity and smaller, one needs to solve the  coupled set of Boltzmann equations for both quark and gluon evolution, with the initial conditions given by the results of the classical-statistical simulations. Again, for a fixed box geometry, this computation has been done~\cite{Blaizot:2014jna}. In an accompanying paper, we will discuss the extension of these results to that of the expanding Glasma~\cite{Tanji:2017suk}. 

The paper is organized as follows. In the next section, we will outline the well known result for photon production from a thermal QGP. Unlike previous computations though, we will study the parametric dependence of the integrated rate on the QCD coupling taking into account the fact that the thermalization time too depends on the coupling. The corresponding estimates for the various stages of Glasma evolution are given in section \ref{sec:BMSS_glasma}. In section \ref{sec:pheno}, we will first discuss how one can constrain the numerical coefficients in the initial temperature and thermalization time from data on charged particle multiplicities. Given these constrained values, we will then compute the relative yields for photon production from the Glasma and the QGP for varying system size and center-of-mass energy.  We will end with a summary and an outlook on future work. 
A formula for the photon production rate in the small-angle approximation is derived in Appendix A.

\section{The thermal photon production rate} \label{sec:thermal}

We will begin this section with a brief recapitulation of well known computations in the literature on the thermal photon production rate. We will subsequently embed this rate in an expanding geometry and discuss the consequences of the parametric dependence of the integrated rate on the BMSS thermalization time relative to the hadronization time. 

\subsection{Recap of known thermal production estimate} \label{subsec:recap}
We will rely on the kinetic expression for the production of on-shell photons with momentum $\bp =(p_x ,p_y, p_z)$ at the space-time point $X=(t,x,y,z)$ from two-to-two scattering~\cite{Kapusta:1991qp,Baier:1991em},
\begin{equation} \label{eq:rate0}
E\frac{dN}{d^4 X d^3p} 
= \frac{1}{2(2\pi)^3} \int_{p_1, p_2,p_3} |\mathcal{M}|^2 (2\pi)^4 \delta^4 (P_1 +P_2 -P_3 -P ) 
f_1 (p_1 )f_2 (p_2 ) \left[ 1\pm f_3 (p_3 ) \right] \, ,
\end{equation}
where
\begin{equation}
\int_p = \int \! \frac{d^3p}{(2\pi)^3 2E_p}
\end{equation}
and $P=(E_p ,\bp )$. 
The squared amplitude $|\mathcal{M} |^2$ should be understood as summed over spins, colors and flavors of all incoming and outgoing particles.
For massless up and down quarks, the amplitude is given in terms of the Mandelstam variables $s=(p_1+p_2)^2$, $t=(p_1-p)^2$ and $u=(p_3-p_1)^2$ as
\begin{equation} \label{eq:anni}
|\mathcal{M}_\text{anni}|^2 = \frac{160}{9} 16\pi^2 \alpha \alpha_s \frac{u^2 +t^2}{ut} \, ,
\end{equation}
for the annihilation process, and
\begin{equation} \label{eq:comp}
|\mathcal{M}_\text{Comp}|^2 = \frac{320}{9} 16\pi^2 \alpha \alpha_s \frac{u^2 +s^2}{-us} \, ,
\end{equation}
for Compton scattering with the electromagnetic coupling constant $\alpha$. 
Because photons are never in equilibrium in the QGP, the formula \eqref{eq:rate0} is valid for both equilibrium and non-equilibrium rates as long as the kinetic description is applicable.

For photon production from a thermal medium, we can simplify subsequent computations considerably by adopting the approximation
\begin{equation}
f_1 (E_1 ) f_2 (E_2 ) \simeq e^{-(E_1 +E_2 )/T} \, . 
\end{equation}
This is a good approximation for high-energy photons with 
$E_1 +E_2 >E \gg T$. 
For $f_3 (E_3)$, one has to keep the Bose-Einstein or the Fermi-Dirac distribution form 
since $E_3$ is not necessarily large. 
Using this simplification, one can arrive at the formula for the thermal photon production rate from two-to-two scattering~\cite{Kapusta:1991qp},
\begin{equation}
E\frac{dN^\text{th}}{d^4 x d^3p} 
= K \frac{5}{9} \frac{\alpha \alpha_s}{2\pi^2} T^2 e^{-E/T} \log \left( \frac{2.912}{g^2} \frac{E}{T}\right) \, .
\label{eq:thermal-rate}
\end{equation}
In the derivation of this expression, the infrared divergence is regulated by the Hard Thermal Loop resummed quark propagator. We have introduced a constant $K$, which we will explain in the next paragraph. 

This rate is leading order (LO) in $\alpha$ and $\alpha_s$. 
However the two-to-three bremsstrahlung and pair annihilation processes, which are naively higher order, contribute at parametrically the same order as the two-to-two processes \cite{Aurenche:1998nw,Aurenche:1999tq,Aurenche:2000gf}. 
The complete LO results\footnote{We note that the next-to-leading $\mathcal{O}(g)$ contribution to the thermal photon rate has been computed; this gives a surprisingly small correction to the LO rate even for  $\alpha_s=0.3$ \cite{Ghiglieri:2013gia}.} which include the Landau-Pomeranchuk-Migdal effect, as well as these collinear enhanced processes, have been derived by Arnold, Moore and Yaffe~\cite{Arnold:2001ms}. 
For $\alpha_s=0.2$, the naive LO rate differs from the complete LO rate by a factor of two in the photon momentum range $2.5 \leq k/T \leq 10$. We introduced the $K$ factor in the above equation to approximately take these effects into account. For the purposes of this study, we will assume $K\simeq 2$ and shall employ the LO formula Eq.~\eqref{eq:thermal-rate} henceforth to estimate the thermal photon yield. 

\subsection{BMSS estimate of the thermal production rate} \label{subsec:BMSS_thermal}
With Eq.~\eqref{eq:thermal-rate} in hand, we shall now make a parametric estimate of the photon yield in the thermal QGP stage based on a simple model for the space-time evolution and the BMSS results on the thermalization time and the initial temperature of the system~\cite{Baier:2000sb}. 

The longitudinally expanding system is conveniently described by the comoving coordinates
$\tau=\sqrt{t^2-z^2}$, $\eta =\text{arctanh} (z/t)$, $\bx_{\!\perp} =(x,y)$ and momentum variables $y_p =\text{arctanh} (p_z/E)$, $\bp_{\!\perp} =(p_x ,p_y)$. 
We will assume for simplicity that the expansion is boost-invariant. The temperature is then a function of proper time,
$T=T(\tau )$. By using the relations
\begin{equation}
d^4 X = \tau d\tau d\eta\, d^2 \xperp \, ,
\end{equation}
and
\begin{equation}
\frac{d^3 p}{E} = dy_p\, d^2 \pperp \, ,
\end{equation}
we can rewrite \eqref{eq:thermal-rate} as
\begin{align}
\frac{1}{S_\perp} \frac{dN^\text{th}}{dy_p d^2 \pperp}
&= K \frac{5}{9} \frac{\alpha \alpha_s}{2\pi^2} \int \! \tau d\tau \int \! d\eta \, 
T^2 e^{-E/T} \log \left( 1+\frac{2.912}{g^2} \frac{E}{T}\right) \notag \\
&\simeq \frac{5}{9} C \frac{\alpha \alpha_s}{2\pi^2}
\int \! \tau d\tau \, T^2 \int \! d\eta \, e^{-E/T} \, , \label{eq:dndydp}
\end{align}
with $S_\perp$ denoting the transverse area. 
In the second line, we have assumed that the $\tau$ and $\eta$ dependence of the logarithmic factor is slower than other factors, and replaced the logarithmic function\footnote{We have added 1 to the argument of the logarithm--as noted in~\cite{Kapusta:1991qp}, adding this constant gives very good agreement between the numerical computation of the rate and the analytic approximation from the $E \gg T$ region where it is justified, down to $E\simeq T$.}
$\log \left( 1+\frac{2.912}{g^2} \frac{E}{T}\right)$ by a numerical factor $\log\left( 1+\frac{2.912}{g^2} \right)$, which is denoted by $C$ after being combined with the $K$ factor:
\begin{equation}
C = K \log \left( 1+\frac{2.912}{g^2} \right) \, .
\label{eq:thermal-constant}
\end{equation}
In a Lorentz covariant description, the photon energy $E$ should be replaced by $p^\mu u_\mu$, 
where $p^\mu$ is the energy-momentum 4-vector and $u_\mu$ is the co-moving 4-velocity,
\begin{equation}
u_\mu = (\cosh \eta ,0,0,\sinh \eta ) \, .
\end{equation}
The $\eta$ integration can be done analytically, 
\begin{align}
\int_{-\infty}^\infty \! d\eta \, e^{-p^\mu u_\mu/T}
&= \int_{-\infty}^\infty \! d\eta \, e^{-p_\perp \cosh (\eta -y_p )/T}
= 2\,K_0 \left( p_\perp /T \right) \, ,
\end{align}
where $K_n (z)$ is the modified Bessel function of the second kind. 
To estimate the QGP photon yield, we need to define $\tau_c$, the time until which any weak coupling estimate of photon production in the QGP may at all be applicable. The most optimistic estimate of $\tau_c$ is the ``hadronization time", the time at which the temperature of the system reaches the crossover temperature,  namely $T=T_c$. We shall adopt this definition henceforth. The photon yield in the QGP can then be expressed as 
\begin{equation}
\frac{1}{S_\perp} \frac{dN^\text{th}}{dy_p d^2 p_\perp} 
= \frac{10}{9} \,C \,\frac{\alpha \alpha_s}{2\pi^2} 
\int_{\tau_\text{th}}^{\tau_c} \! \tau d\tau \, T^2 K_0 \left( p_\perp /T \right) \, .
\label{eq:unintegrated_spectrum0}
\end{equation}
By using the formula
\begin{equation}
\int_0^\infty \! dx\, xK_0 (x) = 1 \, ,
\end{equation}
we can further integrate the yield over the transverse momentum, 
\begin{equation} \label{eq:thermal_yield}
\frac{1}{S_\perp} \frac{dN^\text{th}}{dy_p} 
= \frac{10}{9} C \frac{\alpha \alpha_s}{2\pi^2} 
\int_{\tau_\text{th}}^{\tau_c} \! \tau d\tau \, T^4 \, .
\end{equation}
If we integrate $\pperp$ from $T$ to $+\infty$ since the formula \eqref{eq:thermal-rate} is valid for $\pperp \gtsim T$, 
we obtain a numerical factor of
\begin{equation}
\int_1^\infty \! dx\, xK_0 (x) = K_1 (1) = 0.601... \, .
\end{equation}
This uncertainty can be absorbed by the factor $C$, which anyway possesses order-one uncertainty. 

In order to estimate the thermal photon yield \eqref{eq:thermal_yield}, 
we need a profile for the time dependence of the temperature. 
For simplicity,  we assume a 1+1 dimensional hydrodynamic expansion,
\begin{equation}
\partial_\tau \mathcal{E} = -\frac{\mathcal{E} +P_L}{\tau} 
\end{equation}
with the equation of state
\begin{equation}
P_L =\frac{1}{3} \mathcal{E} \, ,
\end{equation}
where $\mathcal{E}$ is the energy density and $P_L$ is the longitudinal pressure.
These equations can be easily solved to give
\begin{equation}
\mathcal{E} (\tau) = \mathcal{E} (\tau_\text{th}) 
\left( \frac{\tau_\text{th}}{\tau} \right)^{4/3} \, ,
\end{equation}
and from the relation $\mathcal{E} \propto T^4$,  the $\tau$ dependence of the temperature is found to be
\begin{equation} \label{eq:T_timedep}
T (\tau) = T_\text{th}\left( \frac{\tau_\text{th}}{\tau} \right)^{1/3} \, ,
\end{equation}
with $T_\text{th}=T(\tau_\text{th})$ being the temperature at the thermalization time. 
Plugging these expressions into Eq.~\eqref{eq:thermal_yield} leads to the result 
\begin{equation} 
\frac{1}{S_\perp} \frac{dN^\text{th}}{dy_p} 
= \frac{5}{3} C \frac{\alpha \alpha_s}{2\pi^2}  T_\text{th}^4 
\tau_\text{th}^2 \left[ \left( \frac{\tau_c}{\tau_\text{th}}\right)^{2/3} -1\right] \, .
\end{equation}

We now observe that in BMSS \cite{Baier:2000sb,Baier:2002bt}, the time it takes the system to thermalize and  the initial temperature are derived respectively to be 
\begin{equation} \label{eq:BMSS0}
\tau_\text{th} \simeq c_\text{eq} \,\alpha_s^{-13/5} Q_s^{-1}  \hspace{15pt} \text{and} \hspace{15pt}
T_\text{th} \simeq c_T c_\text{eq}\,\alpha_s^{2/5} Q_s \, .
\end{equation}
Here $c_\text{eq}$ is a constant denoting the uncertainty in the BMSS estimate of the time it takes for the Glasma to thermalize -- it can in principle be determined self-consistently in the BMSS framework. As we shall discuss, we will constrain it with data from RHIC and the LHC on hadron multiplicities. There is an additional constant $c_T$ which is needed to determine the initial temperature of the QGP. BMSS determine this number to be, to logarithmic accuracy, $0.16 c$, where $c$ is the gluon liberation constant first discussed in \cite{Mueller:1999fp}. The coefficient $c$ 
is a measure of how efficiently gluons in the wavefunction are released in the collision and can be estimated using boost-invariant classical Yang-Mills simulations of the Glasma~\cite{Krasnitz:2000gz,Krasnitz:2003jw,Lappi:2003bi}; the most sophisticated estimate~\cite{Lappi:2007ku} gives this gluon liberation coefficient to be $c=1.1$. Thus $c_T\simeq 0.18$; however since there are additional logarithmic uncertainties, and $c$ is not known for the full 3+1-D Yang-Mills simulations, we will treat $c_T$ as a constant to be varied within a factor of two of the BMSS value. 

Substituting these expressions for the thermalization time and the initial temperature in our expression for the rate, we obtain
\begin{equation}
\frac{1}{Q_s^2 S_\perp } \frac{dN^\text{th}}{dy_p} 
\simeq \frac{5}{3} c_\text{eq}^6\, c_T^4\, C \frac{\alpha }{2\pi^2}  \alpha_s^{-13/5}
\left[ \left( \frac{\tau_c}{\tau_\text{th}}\right)^{2/3} -1\right] \, ,
\label{eq:int_thermal-intermediate}
\end{equation}
The ratio of the two time scales in the above expression can be obtained from the temporal profile of the temperature \eqref{eq:T_timedep} as
\begin{equation}
\frac{\tau_c}{\tau_\text{th}} = \left( \frac{T_\text{th}}{T_c} \right)^3 \, ,
\label{eq:tauc0}
\end{equation}
where $T_c=154\pm 9$ MeV is the crossover temperature in QCD~\cite{Borsanyi:2010cj,Bazavov:2011nk}; for our study, we will simply take $T_c=154$ MeV. 
Substituting this back in Eq.~\eqref{eq:int_thermal-intermediate}, we obtain our final expression for the thermal yield to be
\begin{align}
\frac{1}{Q_s^2 S_\perp} \frac{dN^\text{th}}{dy_p} 
\simeq \frac{5}{3} c_\text{eq}^6\, c_T^4\, C \frac{\alpha }{2\pi^2} \alpha_s^{-13/5}\,
\left[ c_\text{eq}^2\,c_T^2\,\alpha_s^{4/5}\,\left(\frac{Q_s}{T_c}\right)^2 - 1\right] \, .
\label{eq:int_thermal_estimate-final}
\end{align}

We now summarize the several undetermined constants in our weak coupling expression for the thermal photon yield. The constant $C$, given by Eq.~\eqref{eq:thermal-constant}, is simply the $K$-factor from uncertainities in the thermal rate modulo the logarithmic contribution. As discussed, we will henceforth take $K=2$. 
The constants $c_\text{eq}$ and $c_T$ cannot be completely determined from theory at present. As we will discuss in Sec.~\ref{subsec:constrain}, these constants can be constrained using data on charged hadron multiplicities measured at RHIC and the LHC.  Finally, for $\alpha_s$, we will take the one-loop value assuming it runs with the scale $Q_s$. (This choice also has systematic uncertainties which should be taken into account.) Modulo the stated uncertainties, our result for the thermal photon yield is a function of $Q_s$ alone, which varies both with system size and center-of-mass energy.

\section{Estimate of pre-equilibrium photon production} \label{sec:BMSS_glasma}

In the bottom-up Glasma thermalization scenario of BMSS \cite{Baier:2000sb}, 
the pre-equilibrium Glasma evolution is divided into three temporal stages:
\begin{itemize} \setlength{\itemsep}{-8mm}
\item[(\textbf{i})] $Q_s^{-1} \ll \tau \ll Q_s^{-1} \alpha_s^{-3/2}$ \\
\item[(\textbf{ii})] $Q_s^{-1} \alpha_s^{-3/2} \ll \tau \ll Q_s^{-1} \alpha_s^{-5/2}$ \\
\item[(\textbf{iii})] $Q_s^{-1} \alpha_s^{-5/2} \ll \tau \ll Q_s^{-1} \alpha_s^{-13/5}$ .
\end{itemize} 
In stage (\textbf{i}), the gluons are highly occupied, ranging from an occupancy of $f\sim 1/\alpha_s$ at $\tau\sim 1/Q_s$ to unity at $\tau \sim Q_s^{-1} \alpha_s^{-3/2}$. The occupancy decreases in time as $(Q_s \tau)^{-2/3}$ as  a consequence of the broadening of the longitudinal momentum distribution by elastic scatterings amongst hard gluons. These modify the typical longitudinal momentum from $p_z\sim 1/\tau$ to $p_z \sim  \tau^{-1/3}$. This BMSS prediction is confirmed by the classical-statistical lattice simulations \cite{Berges:2013lsa,Berges:2014yta} which are a good approximation to the real-time dynamics of the theory for $f\gg 1$. 

The quantum kinetic dynamics of the BMSS framework~\cite{Arnold:2002zm} underlies the dynamics of the stages (\textbf{ii}) and (\textbf{iii}), where the occupancy of hard gluons is less than unity. In stage (\textbf{ii}), the soft gluons that are being produced as a result of inelastic scattering dominate screening by providing a larger contribution to the Debye mass relative to that of hard gluons. In stage (\textbf{ii}), the typical longitudinal momentum of hard gluons is $p_z^2 \sim \alpha_s \,Q_s^2$,  which does not depend on time anymore. The anisotropy thus saturates at a value of the ratio of longitudinal to transverse pressure $P_L/P_T \sim \alpha_s$. The multiplicity of soft gluons is however still significantly smaller than those of hard gluons. 

This is no longer the case by the start of stage (\textbf{iii}), with soft gluons dominating the multiplicity for $\tau > Q_s^{-1} \alpha_s^{-5/2}$. Further, $\tau > \tau_{\rm rel}$, the relaxation time of soft gluons; this indicates that the soft gluons have thermalized by then.  The hard gluons however are not thermal, and they thermalize by losing energy to the heat bath of soft gluons through a process which corresponds to the description of jet quenching~\cite{Baier:2000mf}. The infusion of energy into the heat bath raises its temperature (even though the system is expanding) temporarily to saturate finally at the previously quoted temperature of $T_\text{th} = c_T\, c_\text{eq}\,\alpha_s^{2/5} Q_s$ and at the thermalization time of $\tau_\text{th} = c_\text{eq} \, \alpha_s^{-13/5} Q_s^{-1}$. Subsequently, the system undergoes hydrodynamical expansion, with the temperature of the system cooling as $T\sim \tau^{-1/3}$.

We will now estimate photon production from the three stages of the pre-equilibrium evolution of the Glasma. 

\subsection{Glasma stage (\textbf{i})} \label{subsec:glasma1}

To compute the scattering rate in Eq.~\eqref{eq:rate0}, we will explicitly employ the small-angle approximation~\cite{landau1936kinetic,lifshitz1981physical}, which dominates the $2\leftrightarrow 2$ scattering of energetic partons. 
By this approximation, the photon production rate is simplified to
\begin{equation}
E\frac{dN}{d^4X d^3p} = \frac{40}{9\pi^2} \alpha \alpha_s \mathcal{L}  \, f_q (\bp ) 
\int \! \frac{d^3p^\prime}{(2\pi)^3} \frac{1}{p^\prime} \left[ f_g (\bp^\prime ) +f_q (\bp^\prime ) \right] \, ,
\label{eq:saa-rate}
\end{equation}
where $f_g$ and $f_q$ denote the momentum distribution functions of gluons and quarks, respectively. 
The details of this derivation are given in Appendix \ref{sec:small-angle}. 
The symbol $\mathcal{L}$ denotes the so-called Coulomb logarithm, 
\begin{equation}
\mathcal{L} = \int \frac{dq}{q}  \, ,,
\end{equation}
which should be regularized by infrared and ultraviolet cutoffs. In thermal equilibrium, these are respectively the Debye mass $m_D$ and the temperature, giving ${\cal L} \sim \log(1/g)$, which is the origin of the logarithm in Eq.~\eqref{eq:thermal-rate}. 

By integrating \eqref{eq:saa-rate} over $\bp$, one obtains the photon yield per unit rapidity
\begin{equation}
\frac{1}{S_\perp} \frac{dN}{dy_p} = \frac{40}{9\pi^2} (2\pi)^3 \alpha \alpha_s \mathcal{L} \int \! \tau d\tau \, 
I_q (\tau ) \left[ I_g (\tau ) +I_q (\tau ) \right] \, , 
\label{eq:saa_photon}
\end{equation} 
where we have introduced integrals
\begin{equation}
I_{g/q}(\tau ) = \int \! \frac{d^3p}{(2\pi)^3} \, \frac{1}{p} \, f_{g/q} (\bp ,\tau ) \, .
\label{eq:def_Igq}
\end{equation}
Now the problem is reduced to the evaluation of the integrals $I_g$ and $I_q$, which are much simpler than the original multi-dimensional integral in Eq.~\eqref{eq:rate0}. 
In the first stage of the Glasma evolution, $I_q$ in the square bracket is negligible compared to $I_g$.\footnote{%
This indicates that the pair annihilation process is negligible compared with Compton scattering in the gluon dominated medium.}
These integrals are related to the Debye screening mass $m_D$ by the expression $m_D^2 =4\,g^2 (N_c I_g +N_f I_q )$. 

In the first stage of bottom-up thermalization, the Debye mass is dominated by hard gluons whose typical transverse momentum is $\pperp \simeq Q_s$. 
The integral $I_g$ therefore can be approximately related to the number density of hard gluons $n_\text{hard}$ as
\begin{align}
I_g (\tau ) &\simeq \frac{1}{Q_s} \int \! \frac{d^3p}{(2\pi)^3} \, f_g (\bp ,\tau ) \notag \\
&= \frac{1}{Q_s} \frac{n_\text{hard}}{2(N_c^2 -1)} \, ,
\label{eq:Ig_approx}
\end{align} 
where $2\,(N_c^2-1)$ is the degeneracy factor for gluons. 
Since the total number of hard gluons is approximately conserved at this stage, the number density $n_\text{hard}$ decreases in time as $1/\tau$. 
Hence the integral $I_g$ can be represented as
\begin{equation}
I_g (\tau ) 
= \frac{Q_s^2}{\alpha_s} \frac{\kappa_g}{4\pi^2} (Q_s \tau )^{-1} \, ,
\label{eq:Ig}
\end{equation} 
where $\kappa_g$ is a dimensionless constant, which we will fix later.

This functional form of $I_g(\tau)$ is consistent with the following scaling behavior for the gluon distribution, confirmed by classical-statistical simulations~\cite{Berges:2013lsa,Berges:2014yta}:
\begin{equation} 
f_g (\pperp ,p_z ,\tau ) = {1\over \alpha_s}\,(Q_s \tau )^{-2/3} {f}_S \left( \pperp , (Q_s \tau )^{1/3} p_z \right) \, .
\label{eq:scaling-gluon}
\end{equation}
Plugging this scaling form into Eq.~\eqref{eq:def_Igq} leads to the expression, 
\begin{align}
I_g (\tau ) 
&= \frac{1}{\alpha_s} (Q_s \tau )^{-1} \frac{1}{4\pi^2} \int \! \pperp d\pperp \! \int \! d\nu_z \, 
\frac{1}{\sqrt{\pperp^2+(Q_s \tau)^{-2/3} \nu_z^2}} {f}_S \left( \pperp , \nu_z \right) \notag \\
&\simeq \frac{Q_s^2}{\alpha_s} (Q_s \tau )^{-1} \frac{1}{4\pi^2} \int \! \tilde{p}_\perp d\tilde{p}_\perp \! \int \! d\tilde{\nu}_z \, \frac{1}{\tilde{p}_\perp} {f}_S \left( Q_s \tilde{p}_\perp , Q_s \tilde{\nu}_z \right) \, .
\end{align}
If we identify the product of the two dimensionless integrals in the last expression with $\kappa_g$, this expression is equivalent to \eqref{eq:Ig}. 

Since quarks undergo the same scattering processes as gluons, namely small-angle elastic collisions, 
it is natural to assume that the quark occupation number for hard modes with $\pperp \sim Q_s$ decreases with the same power law as that of gluons,
\begin{equation} 
f_q (\pperp ,p_z ,\tau ) = (Q_s \tau )^{-2/3} \tilde{f}_S \left( \pperp , (Q_s \tau )^{1/3} p_z \right) \, .
\label{eq:scaling-quark}
\end{equation}
This functional form of the quark distribution at the scale $\pperp \sim Q_s$ has been confirmed by us in a kinetic treatment {\it a la} BMSS~\cite{Tanji:2017suk}. 
We note that while quark pairs are produced copiously at early times,  the gluon fusion process will continue to produce pairs as the system evolves, albeit at a diminishing rate, with this contribution amenable to a perturbative treatment in a kinetic approach~\cite{Blaizot:2014jna,Ruggieri:2015tsa}. 
In weak coupling, this effect is negligible and the evolution of the quark spectrum is well described by the scaling form \eqref{eq:scaling-quark} \cite{Tanji:2017suk}.
Performing the same computation as in the gluon case, we can express the integral for quarks $I_q$ as
\begin{equation}
I_q (\tau ) = Q_s^2 \frac{\kappa_q}{4\pi^2} (Q_s \tau )^{-1} \, ,
\label{eq:Iq}
\end{equation}
where another constant $\kappa_q$ has been introduced. We note that the factor $1/\alpha_s$ in $I_g$ is absent here.

Substituting the expressions \eqref{eq:Ig} and \eqref{eq:Iq}  into Eq.~\eqref{eq:saa_photon}, we can express the photon yield in stage (\textbf{i}) as
\begin{align}
\frac{1}{Q_s^2 S_\perp} \frac{dN^\text{glasma-i}}{dy_p} 
&= \frac{40}{9\pi} \frac{\alpha}{2\pi^2} \mathcal{L}\, \kappa_g \kappa_q \int_{\tau_0}^{\tau_1} d\tau \, \tau^{-1} \notag \\
&= \frac{40}{9\pi} \frac{\alpha}{2\pi^2} \mathcal{L}\, \kappa_g \kappa_q \, \log \alpha_s^{-3/2} \, .
\label{eq:stagei_yield}
\end{align}
We have used here $\tau_0 \sim Q_s^{-1}$ and $\tau_1 \sim Q_s^{-1} \alpha_s^{-3/2}$. 
This estimate has a systematic uncertainty since we do not fix the numerical coefficients of $\tau_0 \sim Q_s^{-1}$ and $\tau_1 \sim Q_s^{-1} \alpha_s^{-3/2}$. 
However, the uncertainty is small because $\tau_0$ and $\tau_1$ both only appear inside the logarithmic factor. 

We therefore have a good estimate of the non-equilibrium photon yield in the first stage of the Glasma evolution. 
In weak coupling, this yield is parametrically much smaller than the thermal yield in Eq.~\eqref{eq:int_thermal_estimate-final}. However as we noted previously, the thermal rate is sensitive to high powers of the constants $c_T$ and $c_\text{eq}$. We will demonstrate later that, for realistic values of the coupling, the ratio of the two yields will depend strongly on these coefficients.  

The values of the constants $\kappa_g$ and $\kappa_q$, which depend on the normalization of the scaling function $f_S$, can be determined by the following considerations. The number density of hard gluons produced immediately after the heavy-ion collision can be expressed as~\cite{Mueller:1999fp}
\begin{equation}
n_{\rm hard} = c\, \frac{(N_c^2-1) Q_s^3}{4\pi^2 N_c \alpha_s}\,\frac{1}{Q_s\tau} \, ,
\label{eq:nhard}
\end{equation}
where $c$ is the gluon liberation coefficient we discussed previously, with $c = 1.1$ from solutions of the boost-invariant classical Yang-Mills equations~\cite{Lappi:2007ku}.  
Combining this with Eq.~\eqref{eq:Ig_approx} immediately leads to 
\begin{equation}
\kappa_g = \frac{c}{2N_c} \, .
\end{equation}
To compute $\kappa_q$, we assume that the quark number density at $\tau\sim Q_s^{-1}$ is smaller than that of gluons by the factor $\alpha_s$.
Multiplying $\alpha_s$ as well as a factor to convert the degeneracy factor to the gluon number density in Eq.~\eqref{eq:nhard}, we obtain
\begin{equation}
n_\text{quark} = c\,\frac{N_f Q_s^3}{2\pi^2 } \frac{1}{Q_s \tau} \, ,
\end{equation}
from which we find
\begin{equation}
\kappa_q = \frac{c}{2N_c} \, .
\end{equation}

\subsection{Glasma stage (\textbf{ii})} \label{subsec:glasma2}
In this stage, the typical occupancy of hard gluons is less than unity and decreases as $f\sim \alpha_s^{-3/2}/(Q_s \tau )$. While the total particle number is still dominated by hard gluons, the Debye mass is dominated by soft gluons,
and it behaves as $m_D\sim \alpha_s^{3/8} Q_s (Q_s\tau)^{-1/4}$ \cite{Baier:2000sb}. 
Therefore, the time dependence of the integral $I_g$ is different from its dependence in the Glasma first stage, and it can be evaluated as
\begin{equation}
I_g (\tau ) \sim \alpha_s^{-1} m_D^2 \sim \alpha_s^{-1/4} Q_s^2 (Q_s \tau)^{-1/2} \, .
\end{equation}
By assuming this expression agrees with that in the first stage\footnote{Since we have considered only the hard contribution to $I_g$ in stage \textbf{(i)} and the soft one in stage \textbf{(ii)}, the two expressions for $I_g$ do not necessarily agree at $\tau_1$. However this assumption may be appropriate because, as shown in \cite{Baier:2000sb}, the Debye mass receives equal contributions from hard and soft gluons at $\tau_1$.}, given by Eq.~\eqref{eq:Ig}, at $\tau_1=c_1 \, Q_s^{-1} \alpha_s^{-3/2}$, the overall normalization of $I_g$ is identified to be
\begin{equation}
I_g (\tau ) =  \frac{\kappa_g}{4\pi^2} c_1^{-1/2} \alpha_s^{-1/4} \, Q_s^2\,(Q_s \tau)^{-1/2} \, .
\end{equation}
Here we have introduced the numerical coefficient $c_1$ to fix the time scale $\tau \sim Q_s^{-1} \alpha_s^{3/2}$.

For quarks, soft modes never dominate the integral $I_q$. 
Therefore we assume that the time dependence of $I_q$ is the same as that in the first stage\footnote{
This assumption may not be adequate. For chemical equilibration between quarks and gluons, the quark production process is essential. If the total quark number is increasing, the time dependence of $I_q$ should be slower than $\tau^{-1}$ and the photon yield may become larger than the present estimate.}:
\begin{equation}
I_q (\tau ) = Q_s^2 \frac{\kappa_q}{4\pi^2} (Q_s \tau )^{-1} \, .
\end{equation}

Plugging these expressions into \eqref{eq:saa_photon}, we obtain
\begin{align}
\frac{1}{Q_s^2 S_\perp} \frac{dN^\text{glasma-ii}}{dy_p} 
&= \frac{40}{9\pi} \frac{\alpha}{2\pi^2} \mathcal{L} \, \kappa_g \kappa_q \, c_1^{-1/2} \, \alpha_s^{3/4} \, Q_s^{1/2} \int_{\tau_1}^{\tau_2} \! \tau^{-1/2} d\tau \notag \\
&= \frac{80}{9\pi} \frac{\alpha}{2\pi^2} \mathcal{L} \, \kappa_g \kappa_q \left[ \left(\frac{c_2}{c_1}\right)^{1/2} \alpha_s^{-1/2} -1 \right] \, .
\end{align}
To obtain the last expression, we have substituted $\tau_1 = c_1 \, Q_s^{-1} \alpha_s^{-3/2}$ and $\tau_2 = c_2 \, Q_s^{-1} \alpha_s^{-5/2}$, where $c_2$ is another numerical coefficient.
Although the numerical coefficients $c_1$ and $c_2$ are unknown, it is reasonable to assume that they are both $\mathcal{O} (1)$ and their ratio is close to one. Because of the exponent 1/2, the photon yield is less sensitive to the uncertainty of these coefficients compared with the case of  the thermal photon yield \eqref{eq:int_thermal_estimate-final}, in which the numerical coefficients $c_\text{eq}$ and $c_T$ appear in high powers.  
Here we simply assume that $c_1 = c_2$. 
These coefficients can be determined more accurately by real-time lattice simulations and more detailed kinetic theory computations, which are beyond the scope of the present work. 
Finally, we obtain
\begin{align}
\frac{1}{Q_s^2 S_\perp} \frac{dN^\text{glasma-ii}}{dy_p} 
&\simeq \frac{80}{9\pi} \frac{\alpha}{2\pi^2} \mathcal{L} \, \kappa_q \kappa_q \, \left( \alpha_s^{-1/2} -1 \right) \, .
\label{eq:stageii_yield}
\end{align}
In the weak coupling limit, this yield  is much larger than the photon yield from the first stage given in Eq.~\eqref{eq:stagei_yield}. 

\subsection{Glasma stage (\textbf{iii})} \label{subsec:glasma3}
In this final stage of Glasma evolution, the total particle number is dominated by soft gluons, while most of the energy is carried by
a few hard particles. In the bottom-up thermalization scenario,  the soft gluons reach thermal equilibrium first and the hard gluons subsequently lose their energy to the heat bath of soft gluons by a turbulent bremsstrahlung process~\cite{Blaizot:2012fh}.
The temperature of the heat bath increases in time as
\begin{equation}
T= c_T\, \alpha_s^3\, Q_s^2 \,\tau \, .
\end{equation}
Photons produced from this thermal bath can be estimated from the thermal rate \eqref{eq:thermal_yield}
by replacing $\tau_\text{th}$ by $\tau_2 = c_2 \, Q_s^{-1} \alpha_s^{-5/2}$ 
and $\tau_c$ by $\tau_\text{th} = c_\text{eq} \, Q_s^{-1} \alpha_s^{-13/5}$:
\begin{align}
\frac{1}{Q_s^2\,S_\perp}\frac{dN^\text{glasma-iii}}{dy_p} 
&= \frac{10}{9} \frac{\alpha \alpha_s}{2\pi^2} C\, Q_s^{-2} 
\int_{\tau_2}^{\tau_\text{th}} \! \tau T^4 d\tau \notag \\
&\simeq  \frac{5}{27} \frac{\alpha}{2\pi^2}\, c_\text{eq}^6 \, c_T^4\, C\, \left[ \alpha_s^{-13/5} -\left( \frac{c_2}{c_\text{eq}} \right)^6 \alpha_s^{-2} \right] \, .
\label{eq:stageiii_yield}
\end{align}
The second term inside the brackets has a systematic uncertainty since we do not fix the numerical coefficient $c_2$. However, since the temperature increases in this stage of Glasma evolution, the photon yield is dominated by $\tau \sim \tau_\text{th}$. Further, in the limit of weak coupling, the second term is negligible compared to the first term. 
In keeping with our assumption about coefficients multiplying time scales being of the same order, we shall simply henceforth replace the ratio $c_2/c_\text{eq}$ by unity. 

It is interesting to compare the above expression for the Glasma yield in stage (\textbf{iii}) directly with the thermal yield in Eq.~\eqref{eq:int_thermal_estimate-final}. We obtain
\begin{equation}
\frac{dN^\text{glasma-iii}}{dy_p} \bigg/ \frac{dN^\text{th}}{dy_p}  = \frac{1}{9} \frac{1 - \alpha_s^{3/5}}{ c_\text{eq}^2 \, c_T^2 \,\alpha_s^{4/5}\,\left(\frac{Q_s}{T_c}\right)^2 -1} \, .
\end{equation}
In the limit of very weak coupling, which also corresponds to very large values of $Q_s$ (taking into account the running of the coupling with $Q_s$), the ratio of the two yields goes to zero. We will discuss this ratio for realistic values of 
$Q_s$, $c_T$ and $c_\text{eq}$ in Sec.~\ref{subsec:comp}.

One can also have photon emission during the process of quenching the hard quarks to the heat bath. This has been computed previously for a static medium~\cite{Zakharov:2004bi} for the case where only photons are radiated; since the contribution of soft gluon radiation along with that of photons may be significant, this analysis is incomplete. We leave further discussion of this contribution to future work. 

\section{Phenomenological estimates: Thermal versus Glasma photon yields} \label{sec:pheno}

We will now obtain estimates for the thermal photon yield and the corresponding Glasma photon yield,  based on the expressions in the previous sections. The biggest uncertainties are the parameters $c_\text{eq}$ and $c_T$ which appear, with high powers, in the thermal photon yield. 
As we shall now discuss, they are constrained by data from RHIC and the LHC on charged particle multiplicity.

\subsection{Estimates of $c_{\rm eq}$ and $c_T$} \label{subsec:constrain}
In Sec.~\ref{sec:thermal}, we observed that the thermal photon yield is sensitive to the coefficients $c_{\rm eq}$ and $c_T$ that appear in estimates of the thermalization time and the initial temperature at that time. Assuming that the system satisfies nearly ideal hydrodynamic flow conserving entropy subsequently\footnote{It has been shown in Ref.~\cite{Kurkela:2015qoa} that weak-coupling dynamics matches smoothly with hydrodynamic simulations for the couplings we employ in our work. The viscous hydrodynamic simulations employing a temperature-dependent $\eta/s$ produce entropy, however, it is an effect of about 15\% \cite{Schenke:2013dpa,McDonald:2016vlt}, which is part of the systematic uncertainties of our parametric estimates.}, one can use thermodynamic relations and the data on charged particle multiplicities to constrain them \cite{Baier:2002bt,Baier:2011ap}. 

The entropy of hadrons per unit rapidity can be related to the measured multiplicity of charged hadrons as
\begin{equation} \label{eq:S_had}
\frac{dS_\text{hadron}}{d\eta} = k_{S/N} \,\frac{dN_\text{ch}}{d\eta} \, .
\end{equation}
The proportionality constant $k_{S/N}$ can be phenomenologically estimated in several ways as summarized in \cite{Gubser:2008pc}. We shall adopt the value $k_{S/N}=7.2$ that has been extracted from experimental data for particle yields, spectra and source sizes estimated by two-particle interferometry \cite{Pal:2003rz}. 
The entropy of the QGP per unit rapidity at time $\tau$ and temperature $T$ is
\begin{align} \label{eq:S_QGP}
\frac{dS_\text{QGP}}{d\eta} 
&=\frac{2\pi^2}{45} \nu_\text{QGP} \, S_\perp \,\tau T^3 \, ,
\end{align}
where $\nu_\text{QGP} = 2\,(N_c^2-1) + \frac{7}{2}\,N_f\,N_c$ denotes the internal degrees of freedom for the QGP phase; $\nu_{\rm QGP} =37$ for $N_f=2$. 

Equating the entropy in the two phases gives
\begin{equation} \label{eq:constraint_s0}
\frac{74\pi^2}{45} S_\perp \tau T^3 = k_{S/N}\frac{dN_\text{ch}}{d\eta} \, .
\end{equation}
Since entropy is conserved in ideal hydrodynamic evolution, we can estimate the left hand side at any time during this stage.
By substituting $\tau=\tau_\text{th}$ and $T=T_\text{th}$ given by Eq.~\eqref{eq:BMSS0}, we can relate the unknown constants $c_\text{eq}$ and $c_T$ to the charged hadron multiplicity as
\begin{equation} \label{eq:constraint_s1}
c_\text{eq} c_T^{3/4} = \left[ \frac{45}{148 \pi^2} \, k_{S/N} \, \alpha_s^{7/5} \frac{N_\text{part}}{Q_s^2 S_\perp} \frac{2}{N_\text{part}} \frac{dN_\text{ch}}{d\eta} \right]^{1/4} \, .
\end{equation}
We have introduced the number of participants $N_\text{part}$ because experimental data for $\frac{2}{N_\text{part}} \frac{dN_\text{ch}}{d\eta}$ are available as a function of $N_\text{part}$ at RHIC and the LHC \cite{Adler:2004zn,Abelev:2008ab,Aamodt:2010cz,Alver:2010ck}.
In this section, we will express all quantities as a function of $N_\text{part}$. 
For that, we need a model which can relate $Q_s$ and $S_\perp$ to $N_\text{part}$.

\begin{figure}[tb]
 \begin{center}
  \includegraphics[clip,width=8cm]{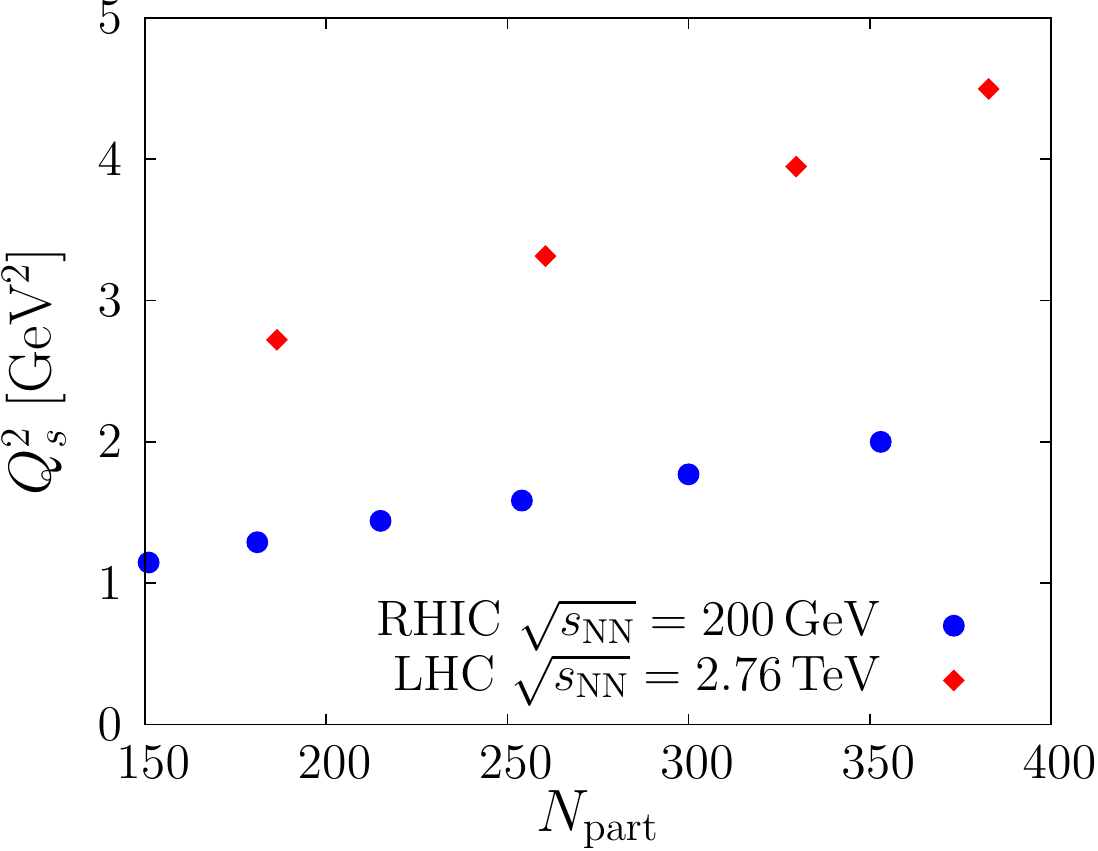} 
 \end{center} \vspace{-20pt}
\caption{The saturation scale squared $Q_s^2$ as a function of $N_\text{part}$ for $\sqrt{s_\text{NN}}=200$ GeV at RHIC and $\sqrt{s_\text{NN}}=2.76$ TeV at the LHC.}
\label{fig:Qs}
\end{figure}

In the IP-Glasma model~\cite{Schenke:2012wb,Schenke:2012fw,Schenke:2013dpa}, the IP-Sat dipole framework of gluon saturation~\cite{Kowalski:2003hm,Rezaeian:2012ji} and the geometrical cross-sections of the Glauber model~\cite{Miller:2007ri} are combined, and $Q_s^2 \,S_\perp \equiv \int dx_T^2 \,Q_s^2(x_T,\sqrt{s})$ can be determined as a function of $N_\text{part}$ and the center-of-mass energy $\sqrt{s}$.\footnote{We thank Prithwish Tribedy for providing us with the values of $Q_s^2 S_\perp$ and $N_\text{part}$ in the IP-Glasma model.}
We further use the Glauber model to compute $S_\perp$ as a function of $N_\text{part}$, and can compute $Q_s^2$ by combining the two results.  
However in the IP-Glasma model, there are two spatially varying saturation scales, one from the projectile and other from the target. At any given spatial position, it is the lower of the two saturation scales that governs the multiplicity of gluons produced locally. Thus the  value of $Q_s^2$ extracted in the aforementioned manner is lower than the value that governs the typical momentum of the produced gluons. 
Thus while we will adopt the $N_\text{part}$ dependence of $Q_s$ provided by the IP-Glasma model, we will account for the harder momentum distribution of produced gluons by multiplying the IP-Glasma $Q_s$ by a numerical factor. 

In Fig.~\ref{fig:Qs}, we plot the values of $Q_s^2$ for $\sqrt{s_\text{NN}}=200$ GeV at RHIC and $\sqrt{s_\text{NN}}=2.76$ TeV at the LHC. 
We have adjusted the overall normalization to obtain a ``reference" $Q_s^2$ at the RHIC most central collision ($N_\text{part}=353$) of 2~GeV$^2$. Absent a first principles determination of the hard scale in the Glasma, this choice of this reference value is somewhat arbitrary with the only consideration being that it is a semi-hard scale. The LHC values are obtained by multiplying the factor $(2.76/0.2)^{0.3}$ to the RHIC values at the same $N_\text{part}$.  This energy dependence is consistent with the empirically observed value for particle multiplicites from RHIC to LHC energies. Because our simple estimate of  photon production is less reliable for peripheral collisions, we do not plot the region  $N_\text{part}<150$. In Sec.~\ref{subsec:comp}, we shall discuss the dependence of the photon yields for the stated $Q_s^2$ reference scale. 

For the running coupling, we employ the one-loop expression, 
\begin{equation}
\alpha_s (Q_s ) = \frac{4\pi}{9 \log \left( Q_s^2 /\Lambda_\text{QCD}^2 \right)} \, ,
\end{equation}
where $\Lambda_\text{QCD}=0.2$ GeV and we have assumed $N_f=3$. 
For the values of $Q_s^2$ plotted in Fig.~\ref{fig:Qs}, the variation of the coupling is rather slow; it varies between $\alpha_s=0.35$--0.42 for the $N_\text{part}$ range for $\sqrt{s_\text{NN}} = 200$ GeV at RHIC and $\alpha_s=0.29$--0.33 for the comparable range for $\sqrt{s_\text{NN}} = 2.76$ TeV at the LHC. 

\begin{figure}[tb]
 \begin{center}
  \includegraphics[clip,width=8cm]{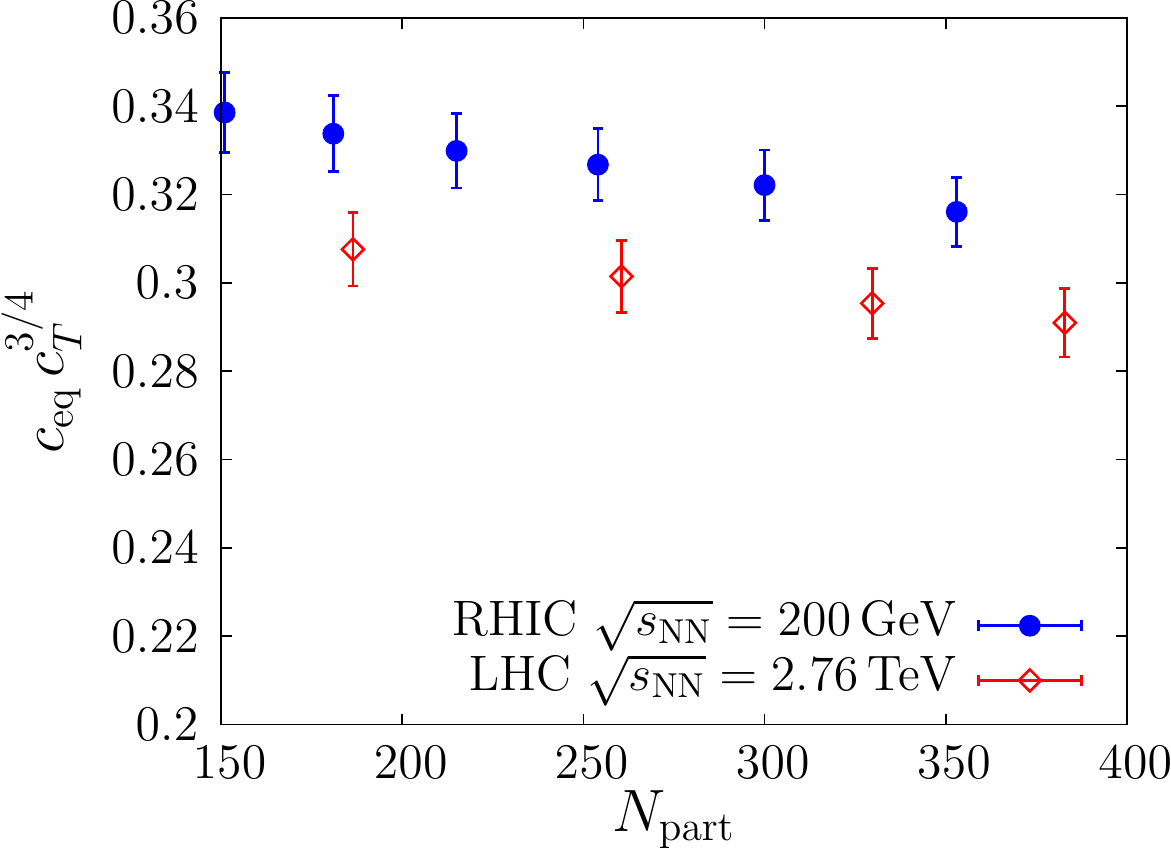} 
 \end{center}
\caption{The numerical coefficient $c_\text{eq}\, c_T^{3/4}$ determined by Eq.~\eqref{eq:constraint_s1}.
The error bars correspond to the systematic errors in the experimental data for the charged hadron multiplicity and those in the Glauber calculation of $S_\perp$.}
\label{fig:coeff}
\end{figure}

For given $Q_s$ and $S_\perp$, we can compute $c_\text{eq}\, c_T^{3/4}$ from the constraint relation in  Eq.~\eqref{eq:constraint_s1}. 
For the values of charged hadron multiplicity, we have used the PHENIX data \cite{Adler:2004zn} and the ALICE data \cite{Aamodt:2010cz}. 
The results are shown in Fig.~\ref{fig:coeff}. 
The values of $c_\text{eq}\, c_T^{3/4}$ are almost independent of $N_\text{part}$ as they should be for our 
analysis to be self consistent. 
For $c_T=0.18$, the value of $c_\text{eq}$ ranges from 1.0 to 1.3. 

Since we have only one constraint equation given by Eq.~\eqref{eq:constraint_s1}, the two coefficients $c_\text{eq}$ and $c_T$ cannot be fixed independently. Only the combination $c_\text{eq}\, c_T^{3/4}$ is fixed\footnote{One may hope to fix the two coefficients independently by further using the measured transverse energy. 
However, the constraint from the transverse energy density is not independent of the hadron multiplicity constraint when the speed of sound is $c_s=1/\sqrt{3}$. 
If $c_s\neq 1/\sqrt{3}$, the two constraints are independent and the two coefficients can be fixed individually. However, the result is very sensitive to the value of $c_s$ and therefore involves a large uncertainty.}.
In the BMSS papers \cite{Baier:2002bt,Baier:2011ap}, $c_T\simeq 0.18$ is estimated\footnote{Note that we have incorporated the gluon liberation constant $c\simeq 1.1$ in our estimate.}  to logarithmic accuracy. To indicate the impact of the uncertainty in this quantity, we will vary it by a factor of two in the range $c_T=0.1$--0.4. 

\subsection{Estimates of  $\tau_{\rm th}$, $T_{\rm th}$, and $\tau_c$ in the bottom-up thermalization scenario} \label{subsec:time-temp}
Before the discussion on the photon yields, it is instructive to show the estimation of the thermalization time $\tau_\text{th}$, the initial temperature $T_\text{th}$, and the hadronization time $\tau_c$ as a function of $N_\text{part}$. Since the coefficients $c_\text{eq}$ and $c_T$ are constrained by the observed hadron multiplicities, we can numerically evaluate these quantities and compare those for RHIC and LHC energies. 
All the results shown in this subsection assume the $Q_s^2$ profile plotted in Fig.~\ref{fig:Qs}. 

\begin{figure}[tb]
 \begin{tabular}{cc}
 \begin{minipage}{0.5\hsize}
  \begin{center}
   \includegraphics[clip,width=7.3cm]{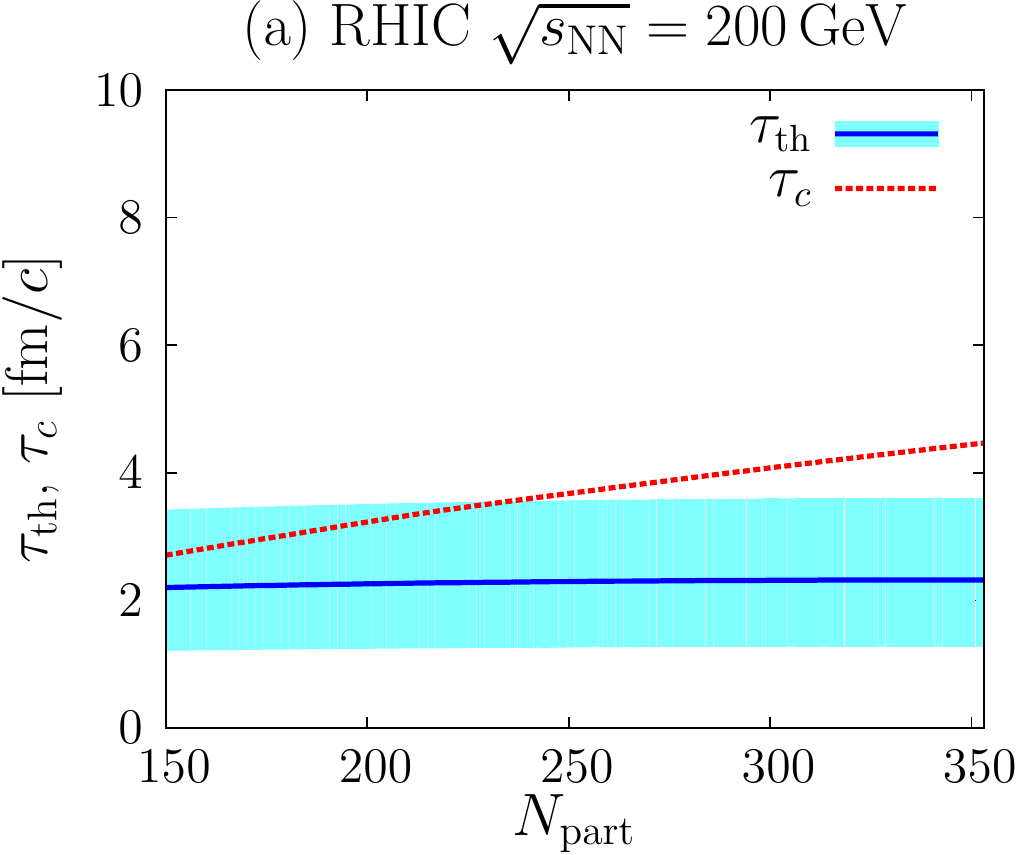} 
  \end{center}
 \end{minipage} &
 \begin{minipage}{0.5\hsize}
  \begin{center}
   \includegraphics[clip,width=7cm]{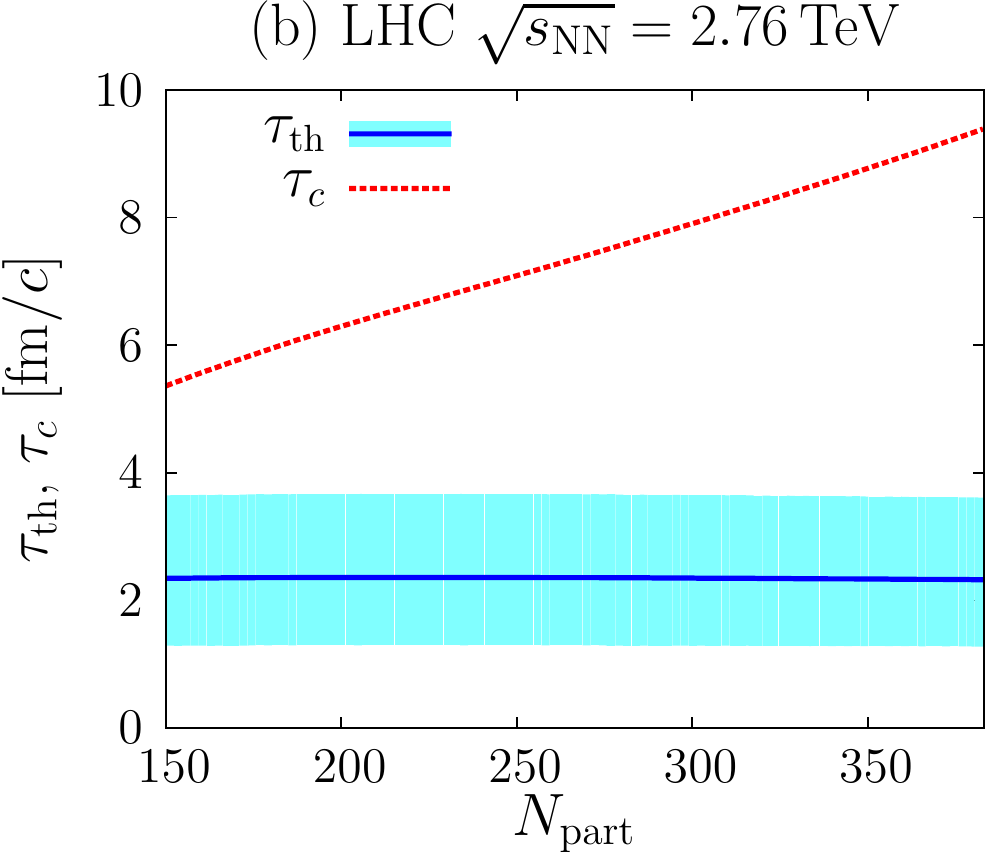} 
  \end{center}
 \end{minipage} 
 \end{tabular}
\caption{The thermalization time $\tau_\text{th}$ and the hadronization time $\tau_c$ as a function of $N_\text{part}$.
Left: RHIC $\sqrt{s_\text{NN}}=200$ GeV. Right: LHC $\sqrt{s_\text{NN}}=2.76$ TeV. 
The color bands denote the uncertainty of $\tau_\text{th}$ corresponding to the variation of $c_T=0.1$--0.4 (top edge of band to bottom edge). The blue solid line corresponds to $c_T=0.18$. }
\label{fig:tauth}
\end{figure}

In Fig.~\ref{fig:tauth}, the thermalization time $\tau_\text{th}=c_\text{eq} \, \alpha_s^{-13/5} Q_s^{-1}$ is plotted as a function of $N_\text{part}$. 
One may expect that $\tau_\text{th}$ is a decreasing function of $N_\text{part}$ because $Q_s$ increases for increasing $N_\text{part}$. However, it is not always true because there is a competition between the factors $\alpha_s^{-13/5}$ and $Q_s^{-1}$. The running coupling $\alpha_s$ is a decreasing function of $N_\text{part}$. Although the variation of the coupling is slow, the factor $\alpha_s^{-13/5}$ varies relatively strongly and it tends to cancel the variation of $Q_s^{-1}$. This is the reason why the plots of $\tau_\text{th}$ are rather flat as a function of $N_\text{part}$ and the values of $\tau_\text{th}$ are similar for RHIC and LHC energies.  
By the same reason, $\tau_\text{th}$ is rather insensitive to the choice of the normalization for the $Q_s^2$ profile in a realistic parameter range. 
In the figure, the uncertainty of $\tau_\text{th}$ corresponding to the variation of $c_T=0.1$--0.4 is expressed by color bands. A larger value of $c_T$ corresponds to a smaller value of $\tau_\text{th}$. 

\begin{figure}[tb]
 \begin{tabular}{cc}
 \begin{minipage}{0.5\hsize}
  \begin{center}
   \includegraphics[clip,width=7.8cm]{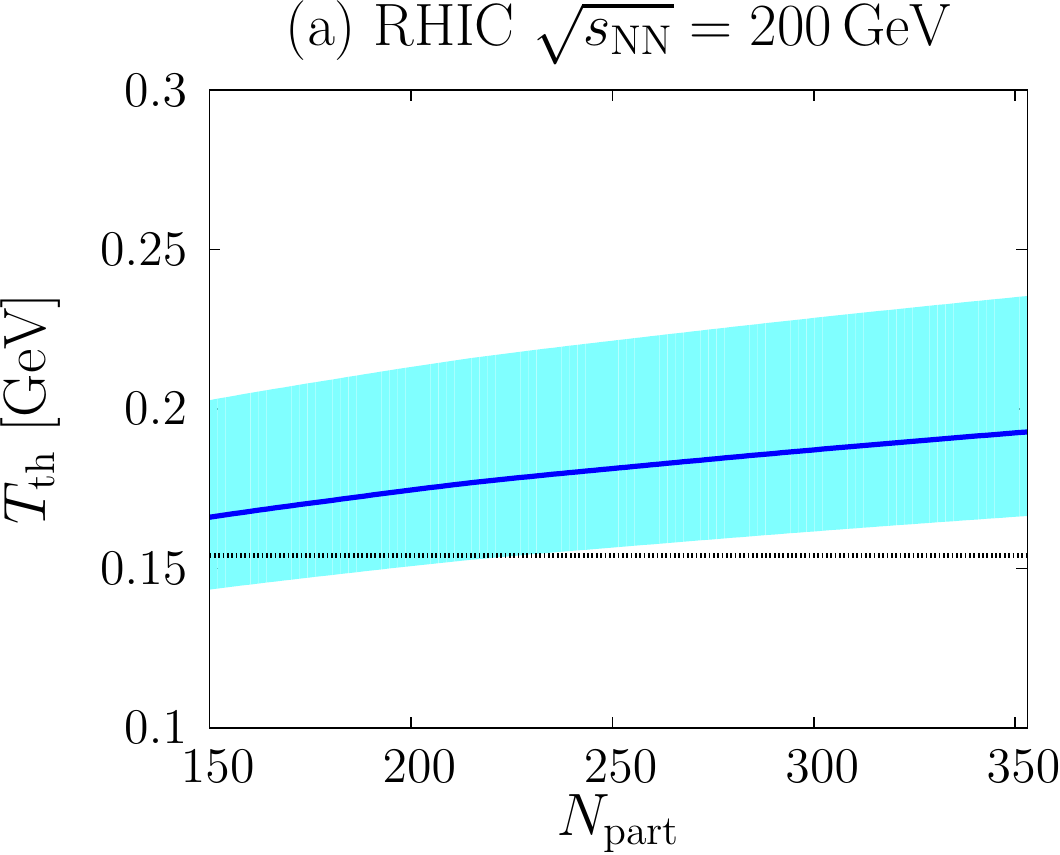} 
  \end{center}
 \end{minipage} &
 \begin{minipage}{0.5\hsize}
  \begin{center}
   \includegraphics[clip,width=7.5cm]{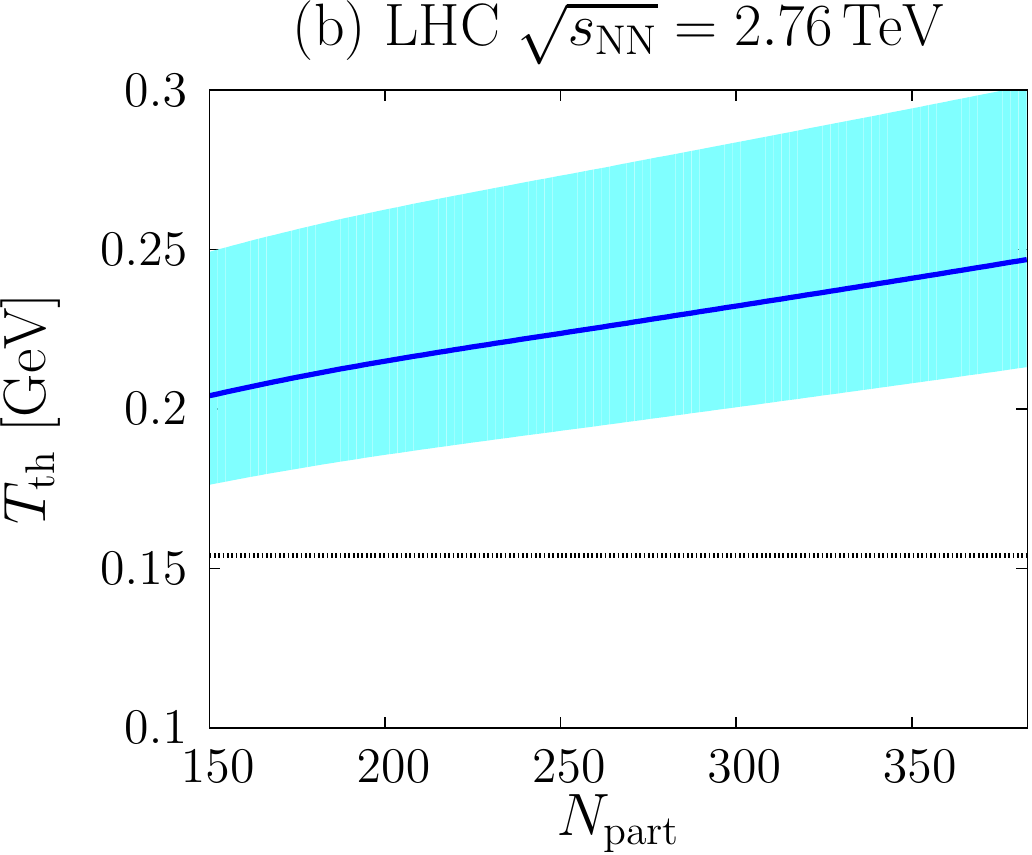} 
  \end{center}
 \end{minipage} 
 \end{tabular}
\caption{The temperature at the time the system thermalizes--plotted as a function of $N_\text{part}$.
Left: RHIC $\sqrt{s_\text{NN}}=200$ GeV. Right: LHC $\sqrt{s_\text{NN}}=2.76$ TeV. 
The color bands represent the uncertainty of $T_\text{th}$ corresponding to the variation of $c_T=0.1$ (bottom edge of band) to $c_T =0.4$ (top edge) with $c_T=0.18$ represented again by the solid blue line. 
The crossover temperature $T_c=154$~MeV is shown as black dashed lines.}
\label{fig:Tth}
\end{figure}

In the same plot, the hadronization time $\tau_c$ is also plotted.
Combining Eqs.~\eqref{eq:tauc0} and \eqref{eq:constraint_s1}, we obtain
\begin{equation}
\tau_c = \frac{45}{74\pi^2}  \, k_{S/N} \frac{1}{S_\perp} \frac{dN_\text{ch}}{d\eta} \frac{1}{T_c^3} \, ,
\label{eq:tauc}
\end{equation}
which is independent of $Q_s$ and $\alpha_s$. 
Therefore, $\tau_c$ is insensitive to the uncertainty of $c_T$ and leads, hence, to the absence of a color band in Fig.~\ref{fig:tauth} for the uncertainty in its value. 
For the highest RHIC energy, the values of $\tau_c$ is only slightly larger than those of $\tau_\text{th}$ for $c_T=0.18$, with the life time of the thermal QGP phase at most 2~fm/$c$ for the most central collisions. For $c_T=0.4$, this is extended to $\simeq 3$ fm/$c$. Clearly, for $N_\text{part} \ltsim 150$, the life time of the thermal QGP phase is quite short -- the Glasma hadronizes out of equilibrium. 
For the LHC energy, $\tau_c$ is larger by about a factor of two and the life time of the QGP phase is significantly longer. 

The temperature at the thermalization time, $T_\text{th}=c_\text{eq}\, c_T\, \alpha_s^{2/5} Q_s$, is plotted in Fig.~\ref{fig:Tth}. 
As in the case of $\tau_\text{th}$, there is a competition between the factors $\alpha_s^{2/5}$ and $Q_s$. However, the exponent of $\alpha_s$ is small here. 
The variation of $Q_s$ wins over that of $\alpha_s^{2/5}$ with the result that $T_\text{th}$ is an increasing function of $N_\text{part}$. 
In the figure, the crossover temperature $T_c=154$~MeV is shown as a black line. 
For the RHIC energy, the thermal QGP initial temperature for $c_T=0.18$ is only slightly larger than $T_c$. 

\subsection{Comparison of thermal and Glasma photon yields} \label{subsec:comp}
We now show the comparison of the photon yields in different stages based on the results obtained in Sec.~\ref{sec:thermal} and \ref{sec:BMSS_glasma}. 
The thermal photon yield is given by Eq.~\eqref{eq:int_thermal_estimate-final}, while the Glasma photon yields in stages \textbf{(i)}, \textbf{(ii)} and \textbf{(iii)} are given in Eqs.~\eqref{eq:stagei_yield}, \eqref{eq:stageii_yield} and \eqref{eq:stageiii_yield}, respectively. 
In the expressions for stages \textbf{(i)} and \textbf{(ii)}, we set the Coulomb logarithm to be $\mathcal{L}= 0.5 \log \left( 1+2.9/g^2\right)$ so that the small-angle approximation reproduces the known formula for the thermal photon yield (see Appendix). 

\subsubsection{Photon production rate} \label{subsubsec:rate}
\begin{figure}[tb]
 \begin{tabular}{cc}
 \begin{minipage}{0.5\hsize}
  \begin{center}
   \includegraphics[clip,width=8cm]{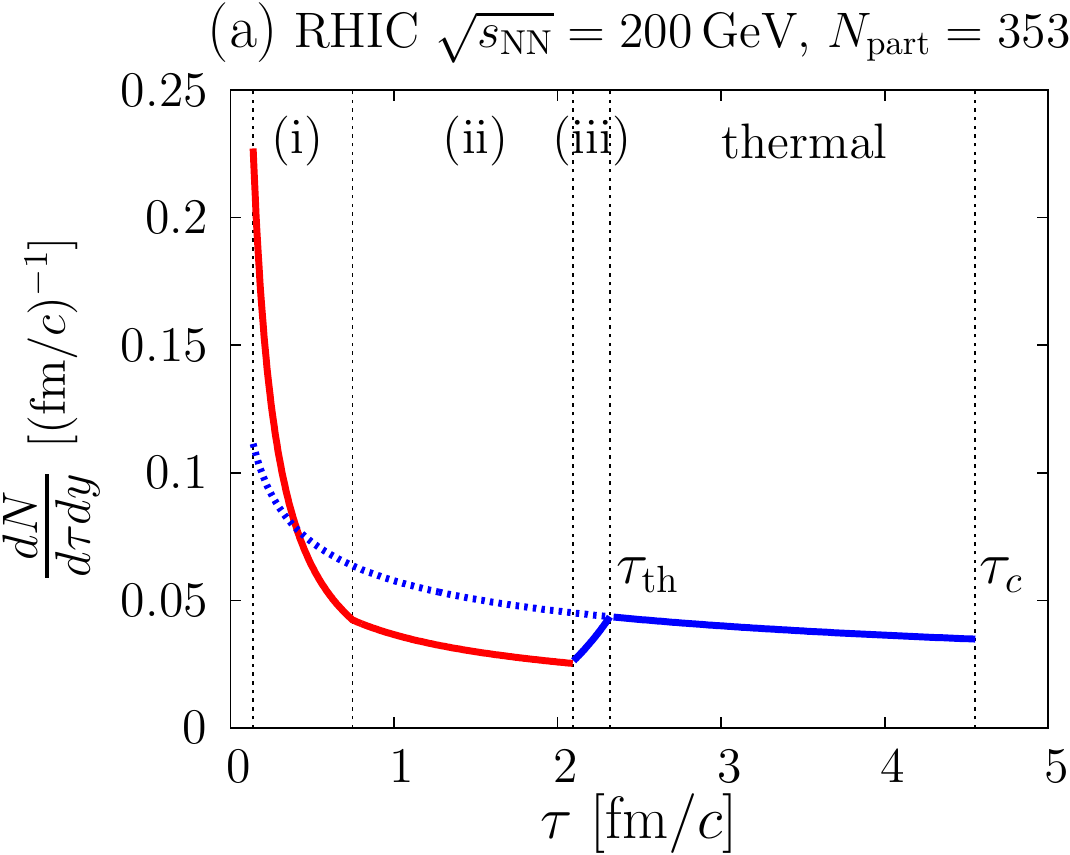} 
  \end{center}
 \end{minipage} &
 \begin{minipage}{0.5\hsize}
  \begin{center}
   \includegraphics[clip,width=8cm]{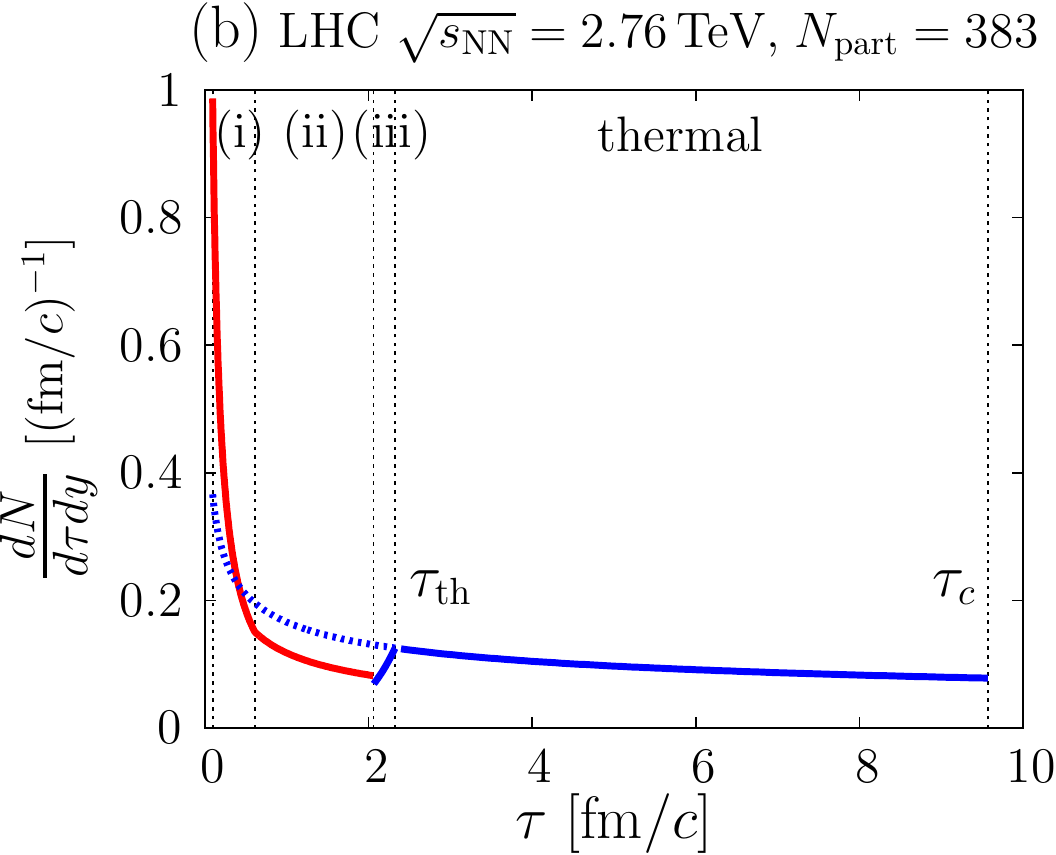} 
  \end{center}
 \end{minipage} 
 \end{tabular}
\caption{Photon production rate as a function of time. 
The results shown are for the most central collisions (centrality 0--5\%); (a) $N_\text{part}=353$ for RHIC, (b) $N_\text{part}=383$ for the LHC.
The red lines represent the nonequilibrium production rate in the Glasma stages \textbf{(i)} and \textbf{(ii)}, while the blue solid lines denote the thermal production rate in the Glasma stage \textbf{(iii)} and the thermal QGP phase. 
The blue dashed lines denote the thermal production rate extended to early times, that we shall call ``early-hydro''.
The vertical black dashed lines separate the different stages of the time evolution.
The value of the coefficient $c_T$ is fixed to 0.18. }
\label{fig:rate}
\end{figure}

To gain insight into photon production in the different stages, we firstly plot, in Fig.~\ref{fig:rate}, the photon production rate $dN/d\tau dy_p$ for the most central collisions as a function of time. 
For simplicity, we fix the value $c_T$ to be 0.18. 
For the Glasma stages \textbf{(i)} and \textbf{(ii)}, we plot the the nonequilibrium production rate $\propto \tau I_g (\tau) I_q (\tau)$ obtained from Eq.~\eqref{eq:saa_photon}  (red lines). For the Glasma stage \textbf{(iii)} and the thermal QGP phase, we show the thermal production rate $\propto \tau T^4$ (blue solid lines). 
In the stage \textbf{(i)}, the rate decreases as $\tau^{-1}$, while in the stage \textbf{(ii)}, the decrease slows down to $\tau^{-1/2}$. Since we do not consider all the possible sources of the photon production, the lines are disconnected between stages \textbf{(ii)} and \textbf{(iii)}.\footnote{In Fig.~\ref{fig:rate}, the red and the blue line seem to be connected. However, this is accidental for the current choice of the $Q_s^2$ profile. As we will discuss below, the Glasma yield strongly depends on the value of $Q_s^2$, while the thermal yield is insensitive to it.} In the stage \textbf{(iii)}, the rate increases as $\tau^5$ since the temperature increases linearly in time. However, the contribution from the stage \textbf{(iii)} is relatively small because the time duration of this stage is short. In the thermal stage, the rate decreases as $\tau^{-1/3}$. 
The area under these lines corresponds to the total photon yield $dN/dy_p$. 

In current hydro simulations for thermal photon production, the bottom-up thermalization scenario is not implemented and hydro modeling is sometimes initialized at early times. For example, in Ref.~\cite{Paquet:2015lta}, the classical Yang-Mills equation is solved with the IP-Glasma initial conditions up to $\tau_0 =0.4$~fm$/c$ and the system evolution is instantaneously switched to hydrodynamic evolution of a thermal QGP. 
This situation corresponds to, within our simple model, extending the thermal lines to early times as represented by blue dashed lines in Fig.~\ref{fig:rate}. 
For comparison, in addition to the bottom-up thermalization scenario, we will consider such hydro scenario extended to the early time, and call the extended hydro stage before $\tau_\text{th}$ ``early-hydro''.

\pagebreak 
In the following, we will compare three contributions: 
\begin{itemize}
\setlength\itemsep{-1mm}
\item the Glasma contribution in $\tau_0 <\tau <\tau_\text{th}$ 
\item the thermal contribution in $\tau_\text{th} <\tau <\tau_c$ 
\item the early-hydro contribution in $\tau_0 <\tau <\tau_\text{th}$. 
\end{itemize}
In the bottom-up thermalization scenario, the total photon yield until the hadronization time is given by the sum of the Glasma contribution and the thermal contribution, while in the hydro scenario that assumes early thermalization, the total yield is the sum of the early-hydro and the thermal contribution. 

\subsubsection{Dependence on $Q_s$} \label{subsubsec:Qs-dep}
Thus far, we have fixed the profile of $Q_s$, as shown in Fig.~\ref{fig:Qs}, by choosing the reference value (the value at the RHIC most central collision) to be 1.4 GeV. 
If one has a complete description of the space-time evolution in heavy-ion collisions, the effective value of $Q_s$ would be fixed for given hadron multiplicity and collision energy. 
However,  as noted previously, because of our lack of  knowledge, the reference value of $Q_s$ cannot be specified within our framework. 
We will therefore treat the reference $Q_s$ as a free parameter and investigate the dependence of the photon yield on it for given values of the measured charged hadron multiplicity.

\begin{figure}[tb]
 \begin{tabular}{cc}
 \begin{minipage}{0.5\hsize}
  \begin{center}
   \includegraphics[clip,width=7.8cm]{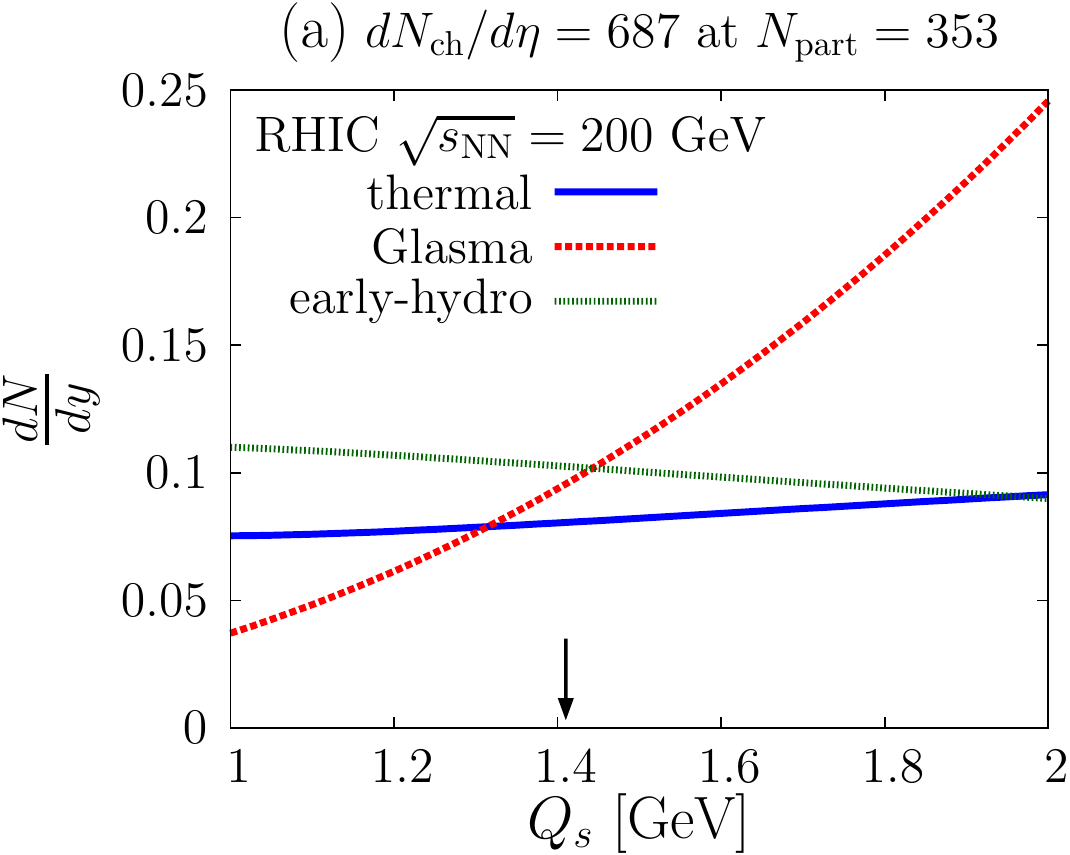} 
  \end{center}
 \end{minipage} &
 \begin{minipage}{0.5\hsize}
  \begin{center}
   \includegraphics[clip,width=7.5cm]{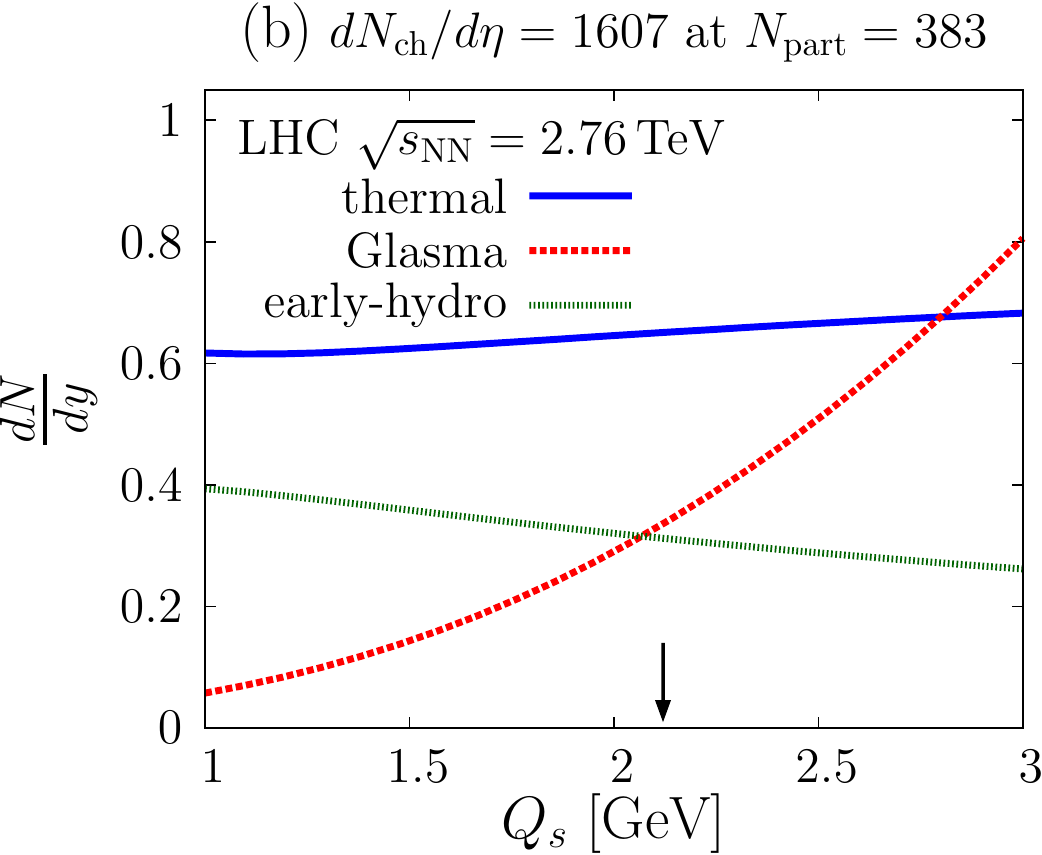} 
  \end{center}
 \end{minipage} 
 \end{tabular}
\caption{Dependence of the photon yield on $Q_s$. For given values of the measured charged hadron multiplicity, we vary the value of $Q_s$.
The number of participants and the transverse area are fixed to those in the most central collisions (centrality 0--5\%); (a) $N_\text{part}=353$ and $S_\perp =140$ fm$^2$ for RHIC, (b) $N_\text{part}=383$ and $S_\perp =156$ fm$^2$ for the LHC. 
The coefficient $c_T$ is fixed to 0.18. 
The values of $Q_s$ assumed in the profile shown in Fig.~\ref{fig:Qs} are indicated by black arrows.}
\label{fig:Qs-dep1}
\end{figure}

We plot in Fig.~\ref{fig:Qs-dep1} the bottom-up thermal photon yield (blue solid line) and the Glasma photon yield (red dashed line) as a function of $Q_s$. For comparison, the early-hydro photon yield (green dotted line) is also shown in the figure. 
The values of $N_\text{part}$ and $S_\perp$ are fixed to those in the most central collisions (centrality 0--5\%) and we have used the corresponding experimental data for the charged hadron multiplicity to find the value of the coefficient $c_\text{eq}$. 
For simplicity, we have set $c_T=0.18$. 
We note that the total photon yield until the hadronization time within the bottom-up thermalization scenario corresponds to the sum of the blue and red lines, while the photon yield in the hydro scenario extended to the early time is given by the sum of the blue and green lines. The respective net contributions will be compared later in Fig.~\ref{fig:comp_early}.

The bottom-up thermal photon yield is not strongly dependent on $Q_s$. This can be accounted for by rewriting Eq.~\eqref{eq:int_thermal-intermediate} as
\begin{equation}
\frac{dN^\text{th}}{dy_p} 
= \frac{5}{3} \left( \frac{45}{74 \pi^2} \right)^{4/3} \, C \frac{\alpha \alpha_s}{2\pi^2}  \left( k_{S/N}\, \frac{dN_\text{ch}}{d\eta} \right)^{4/3} S_\perp^{-1/3} \left( \tau_c^{2/3} -\tau_\text{th}^{2/3} \right) \, ,
\label{eq:thermal_yield_Nch} 
\end{equation}
where we have used Eq.~\eqref{eq:constraint_s1}. 
In this expression, most of the factors are independent of $Q_s$. 
As we discussed previously, $\tau_c$ is completely fixed by the measured hadron multiplicity in our model (see Eq.~\eqref{eq:tauc}). 
The coupling $\alpha_s$ depends on $Q_s$ only logarithmically. 
Only $\tau_\text{th}$ has a power law dependence on $Q_s$. However, this dependence is weak due to the competition between the factors $\alpha_s^{13/5}$ and $Q_s^{-1}$, as noted previously in the context of Fig.~\ref{fig:tauth}, thereby explaining the insensitivity of the thermal photon yield to the normalization of $Q_s$. 
The early-hydro photon yield is obtained from \eqref{eq:thermal_yield_Nch} by replacing $\tau_\text{th}$ by $\tau_0$ and $\tau_c$ by $\tau_\text{th}$. Therefore, it is also insensitive to the value of $Q_s$. 

In contrast, the Glasma photon yield has a strong dependence on $Q_s$. 
This can be easily understood from the expressions in Eq.~\eqref{eq:stagei_yield} and Eq.~\eqref{eq:stageii_yield}, which are the dominant contributions to the Glasma photon yield. 
In these expressions for $\frac{1}{Q_s^2 S_\perp} \frac{dN}{dy}$, the right hand sides are independent of $Q_s$ except for the weak dependence through the coupling. Therefore, $dN/dy$ is approximately proportional to $Q_s^2$, which is consistent with geometrical scaling of direct-photon production discussed in \cite{Klein-Bosing:2014uaa}.
These results indicate that the preequilibrium Glasma photon production can dominate over the thermal one depending on the value of $Q_s$. While the Glasma dominance is pronounced at RHIC, it also begins to dominate at the LHC for $Q_s\simeq 3$ GeV. 

\subsubsection{Dependence on $N_{\rm part}$} \label{subsubsec:Npart-dep}
We now turn to the study of the dependence of the photon yield on $N_\text{part}$ by fixing the profile for $Q_s^2$ as shown in Fig.~\ref{fig:Qs}. The corresponding values of $Q_s$ are indicated in Fig.~\ref{fig:Qs-dep1} by black arrows.  

\begin{figure}[tb]
 \begin{tabular}{cc}
 \begin{minipage}{0.5\hsize}
  \begin{center}
   \includegraphics[clip,width=7.9cm]{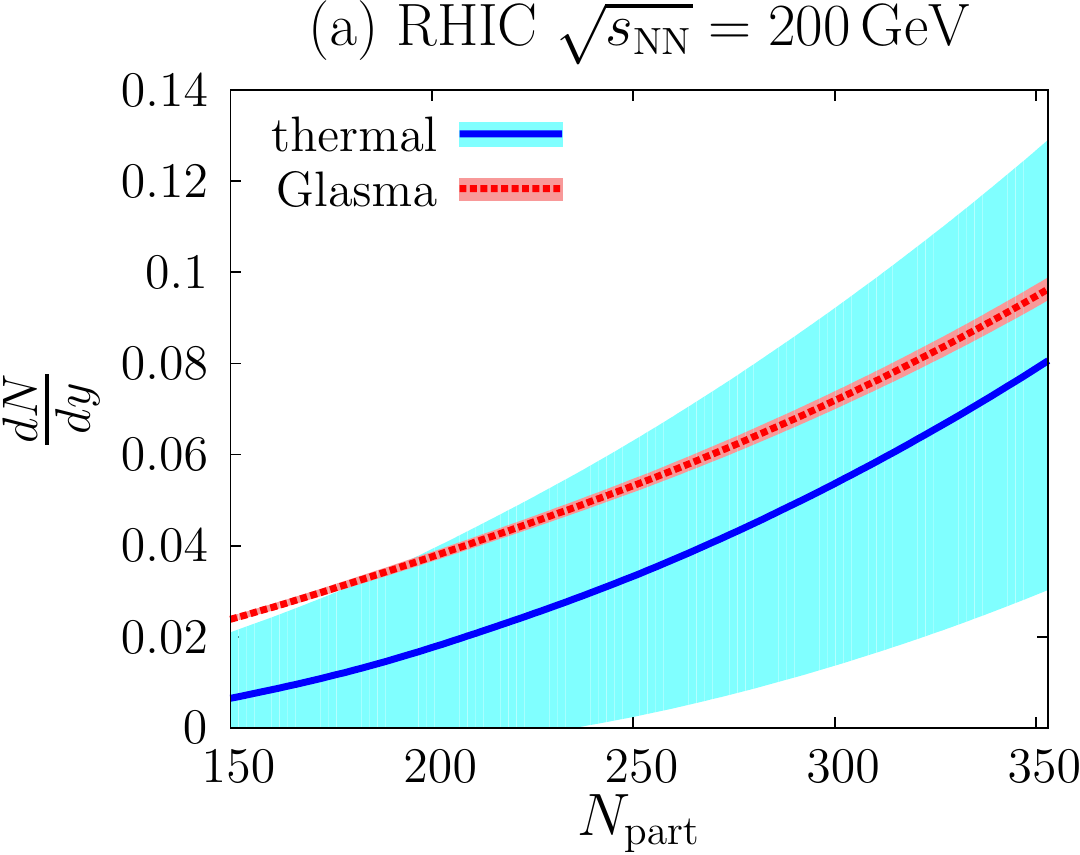} 
  \end{center}
 \end{minipage} &
 \begin{minipage}{0.5\hsize}
  \begin{center}
   \includegraphics[clip,width=7.6cm]{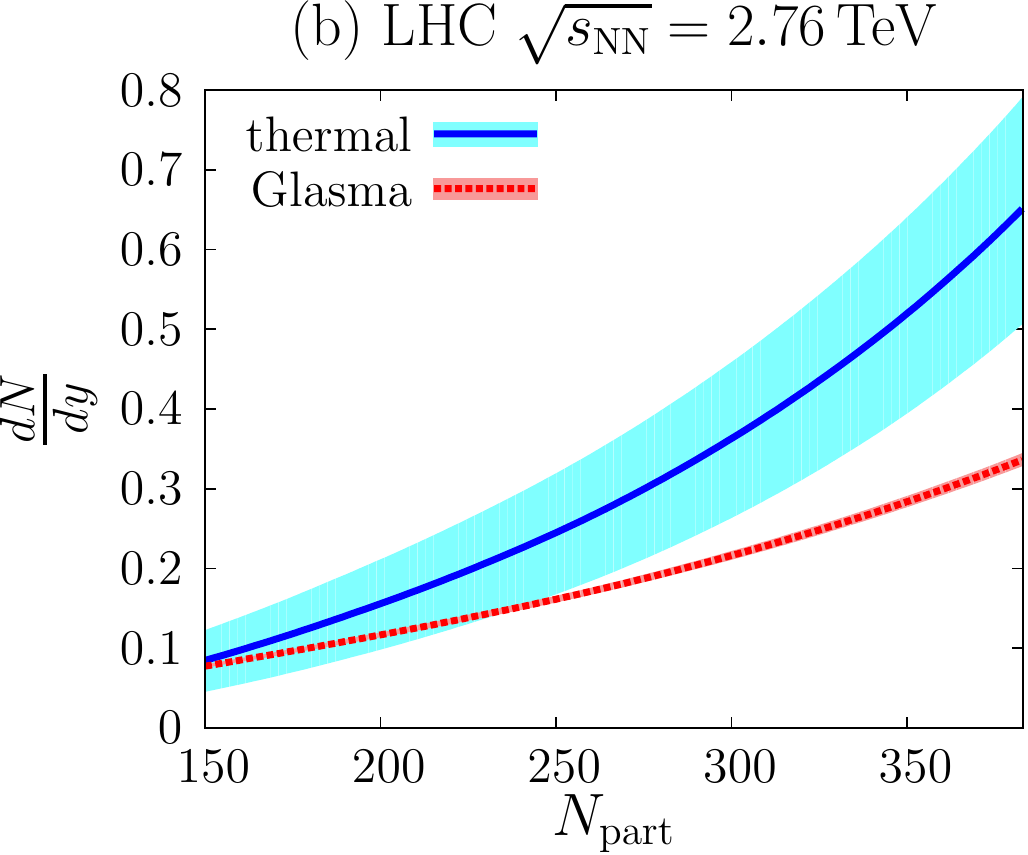} 
  \end{center}
 \end{minipage} 
 \end{tabular}
\caption{Comparison of the thermal photon yield and the preequilibrium Glasma photon yield as a function of $N_\text{part}$.
Left: RHIC $\sqrt{s_\text{NN}}=200$ GeV. Right: LHC $\sqrt{s_\text{NN}}=2.76$ TeV. 
The color bands denote the uncertainty of $\tau_\text{th}$ corresponding to the variation of $c_T=0.1$ (bottom edge of blue band, top edge of red band) to $c_T=0.4$ (top of blue, bottom of red). 
The profiles for $Q_s^2$ are assumed to be those in Fig.~\ref{fig:Qs}.}
\label{fig:photon_comp1}
\end{figure}

In Fig.~\ref{fig:photon_comp1}, we plot the thermal photon yield and the Glasma photon yield as a function of $N_\text{part}$. 
Both of the contributions have the uncertainty associated with the undetermined constant $c_T$. As for previous figures in Sec.~\ref{subsec:time-temp}, this is expressed by color bands corresponding to the range $c_T =0.1-0.4$, with the result for the central value $c_T=0.18$ denoted by the solid curve. 

The RHIC thermal photon yield has a large relative uncertainty. This is because the values of $\tau_c$ and $\tau_\text{th}$ are very close and the life time of the QGP can be short for a small $c_T$ as discussed previously in the context of Fig.~\ref{fig:tauth}. On the other hand, the Glasma photon yield has a small uncertainty because the stage \textbf{(i)} and \textbf{(ii)} contributions are independent of $c_T$.\footnote{We note that, however, the Glasma contributions involve the other systematic uncertainties, e.g. estimation of the functions $I_g$, $I_q$ and the Coulomb logarithm $\mathcal{L}$, which are not reflected in the figures.}
With the exception of the largest values of $c_T$, we observe that, in the bottom-up framework, the Glasma contribution is larger than the thermal QGP contribution to the photon yield for the highest RHIC energy. In particular, for off-central collisions, the relative contribution from Glasma becomes more important. 

For the LHC energy of $2.76$ TeV, the Glasma photon yield for the most central collisions ranges from $40$--$60$\% of the thermal QGP contribution. For $N_\text{part} =150$ it is comparable to the central QGP value, and dominates for more peripheral collisions. 

As we have seen in Fig.~\ref{fig:Qs-dep1}, the Glasma photon yield strongly depends on the value of $Q_s^2$. Therefore the quantitative comparison between the Glasma and the thermal yields also depends on the normalization of the $Q_s^2$ profile. Nevertheless,  for off-central collisions, the qualitative observation that the Glasma contribution is relatively more important is true for any values of $Q_s^2$. 

\begin{figure}[tb]
 \begin{tabular}{cc}
 \begin{minipage}{0.5\hsize}
  \begin{center}
   \includegraphics[clip,width=8.0cm]{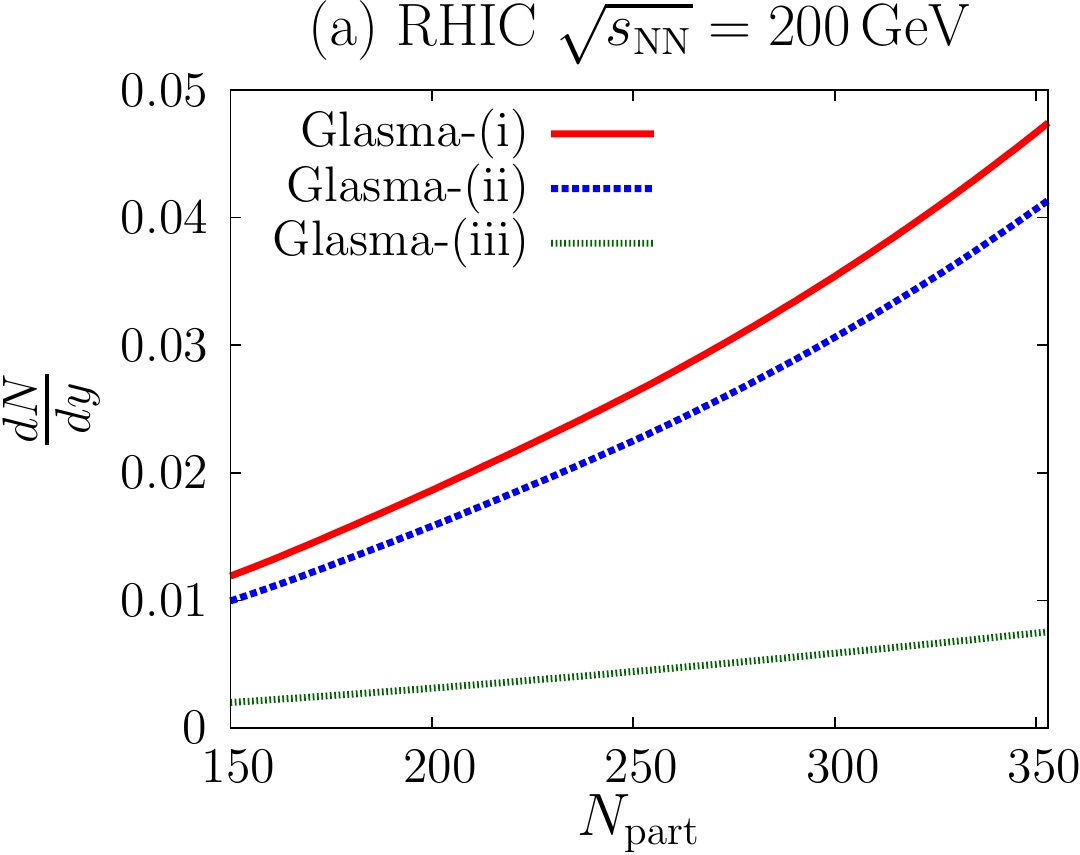} 
  \end{center}
 \end{minipage} &
 \begin{minipage}{0.5\hsize}
  \begin{center}
   \includegraphics[clip,width=7.8cm]{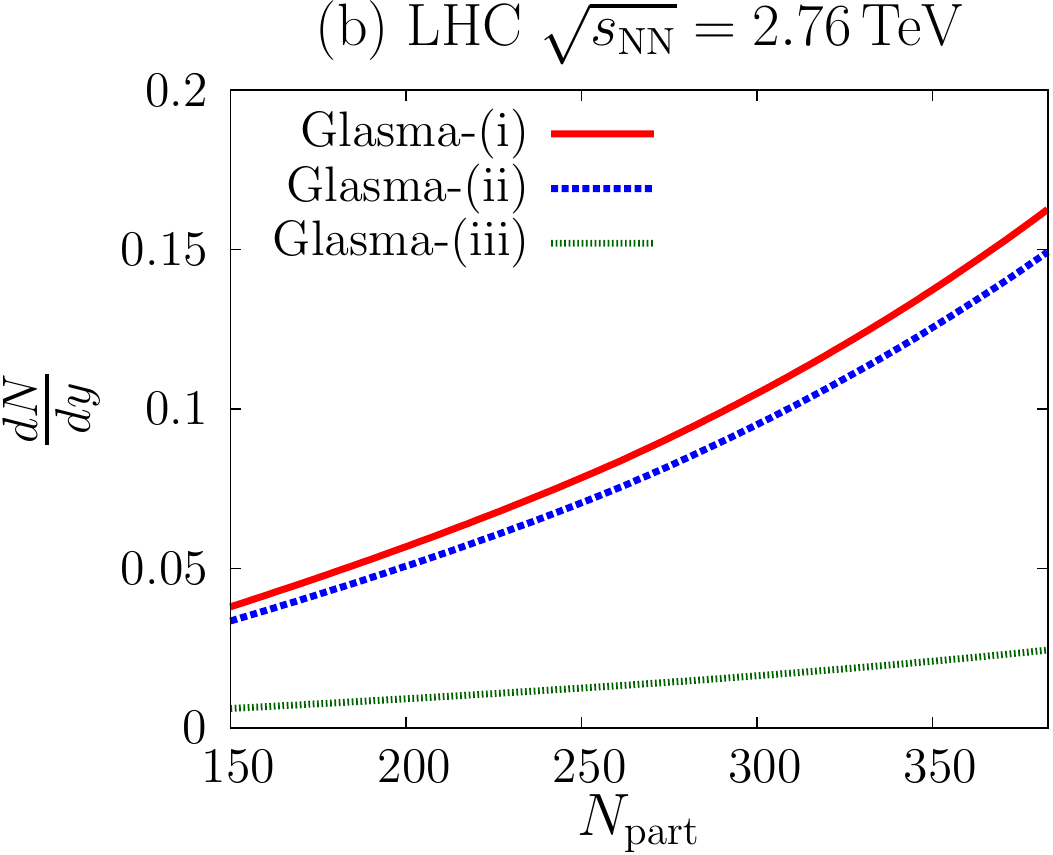} 
  \end{center}
 \end{minipage} 
 \end{tabular}
\caption{Comparison of the photon yields in the three stages of the bottom-up thermalization. 
Left: RHIC $\sqrt{s_\text{NN}}=200$ GeV. Right: LHC $\sqrt{s_\text{NN}}=2.76$ TeV. 
The value of the coefficient $c_T$ is fixed to 0.18. 
The profiles for $Q_s^2$ are assumed to be those in Fig.~\ref{fig:Qs}.}
\label{fig:photon_comp2}
\end{figure}

In Fig.~\ref{fig:photon_comp2}, the photon yields in the Glasma stages \textbf{(i)}, \textbf{(ii)} and \textbf{(iii)} are plotted separately. For the stage \textbf{(iii)} yield, we fix $c_T=0.18$ since the uncertainty is anyway not significant. 
In stage \textbf{(i)}, photons are produced mainly by the scattering among hard gluons and hard quarks, whose transverse momenta are $\sim Q_s$.  In stage \textbf{(ii)}, we have estimated the photon yield by taking into accouunt the scattering between soft gluons and hard quarks. 
In stage \textbf{(iii)}, we evaluated the emission of photons from the bath of soft gluons and quarks. 
Both of the RHIC and the LHC plots display the same systematics; stages \textbf{(i)} and \textbf{(ii)} give similar contributions to the photon yield and the stage \textbf{(iii)} yield is smaller than the yield from the other two stages. 
As discussed in Sec.~\ref{subsec:glasma3}, however, the photon production by the quenching processes of the hard quarks and gluons is not considered in the present study. Such mini-jet photon production processes may have an important contribution in stage \textbf{(iii)}. 

\begin{figure}[tb]
 \begin{tabular}{cc}
 \begin{minipage}{0.5\hsize}
  \begin{center}
   \includegraphics[clip,width=7.9cm]{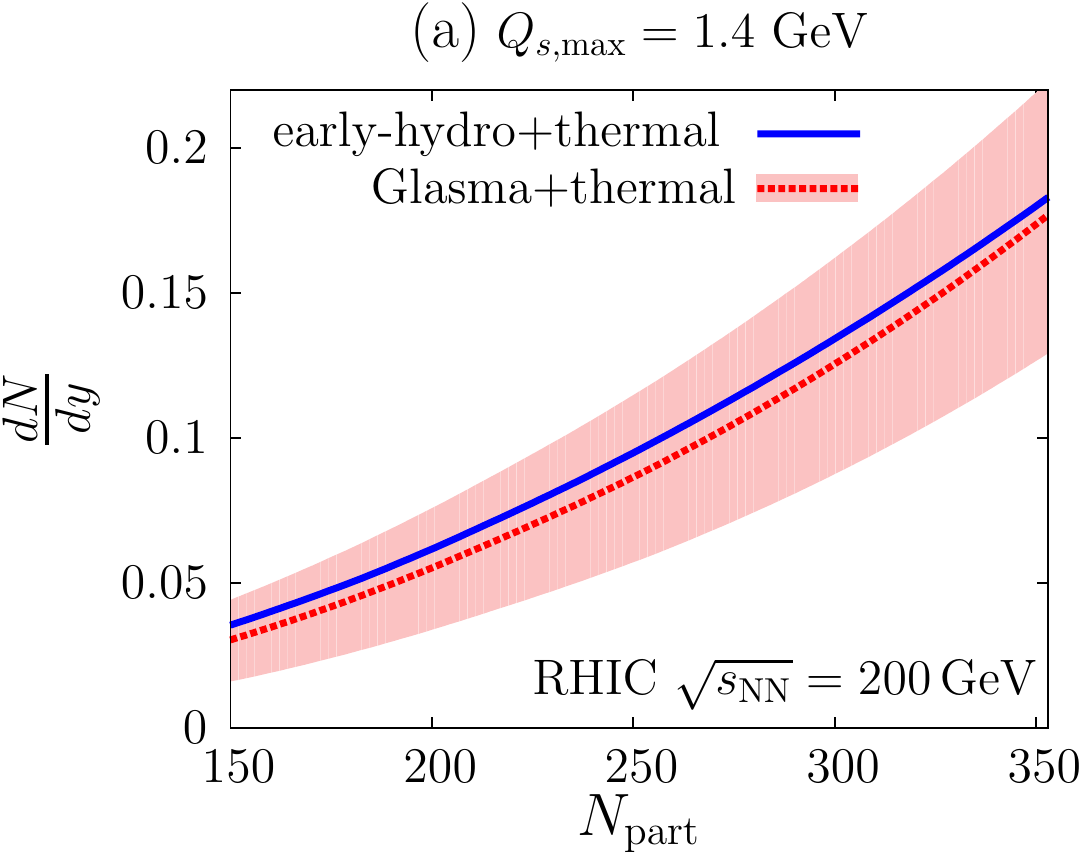} 
  \end{center}
 \end{minipage} &
 \begin{minipage}{0.5\hsize}
  \begin{center}
   \includegraphics[clip,width=7.5cm]{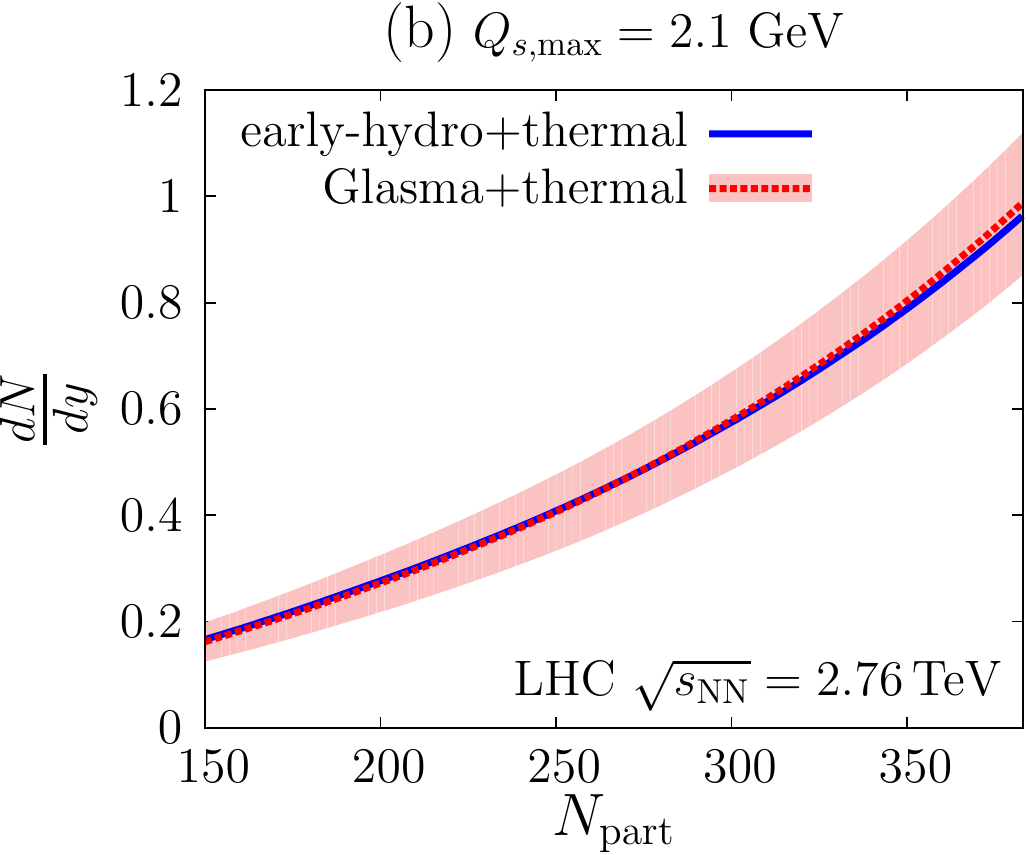} 
  \end{center}
 \end{minipage} \\ \ & \ \\
 \begin{minipage}{0.5\hsize}
  \begin{center}
   \includegraphics[clip,width=7.9cm]{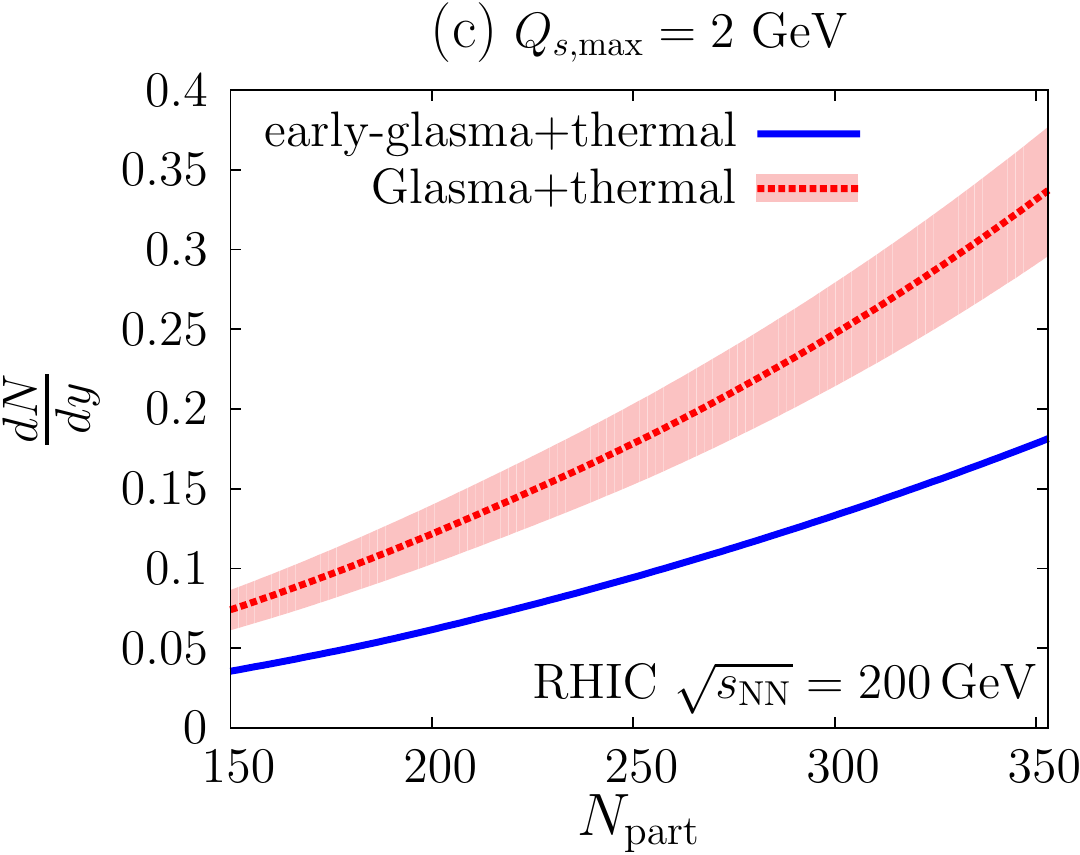} 
  \end{center}
 \end{minipage} &
 \begin{minipage}{0.5\hsize}
  \begin{center}
   \includegraphics[clip,width=7.5cm]{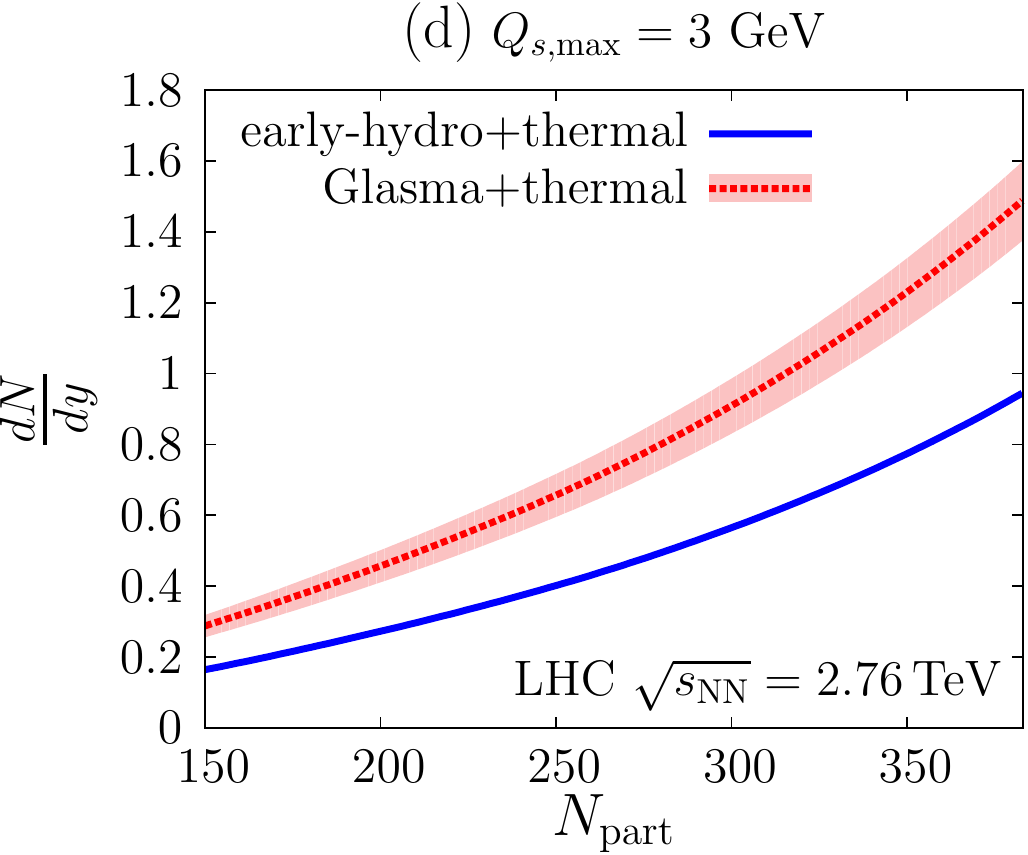} 
  \end{center}
 \end{minipage} 
 \end{tabular}
\caption{Comparison of the photon yields in the bottom-up thermalization scenario and in the hydro scenario that assumes early thermalization. 
Top: The normalization of $Q_s$ is chosen such that the value at the RHIC most central collision is 1.4~GeV (values shown in Fig.~\ref{fig:Qs} and indicated by black arrows in Fig.~\ref{fig:Qs-dep1}). Bottom: The  normalization of $Q_s$ is such that the value at the RHIC most central collision is 2~GeV (the maximum $Q_s$ values shown in Fig.~\ref{fig:Qs-dep1}).
The red color bands denote the uncertainty for the bottom-up themalization yield corresponding to the variation of $c_T=0.1$--0.4 (bottom edge of band to top edge).   
The red dashed lines correspond to $c_T=0.18$. }
\label{fig:comp_early}
\end{figure}

As we have already noted before, the total photon yield until the hadronization time, within the bottom-up thermalization scenario, is given by the sum of the Glasma and the thermal contributions. On the other hand, the photon yield in the hydro scenario that assumes early thermalization corresponds to the sum of the early-hydro contribution (for $\tau_0 <\tau <\tau_\text{th}$) and the thermal contribution (for $\tau_\text{th} <\tau <\tau_c$).
The photon yields in these two different scenarios are compared in Fig.~\ref{fig:comp_early}. 
The contributions that were separately shown in Fig.~\ref{fig:Qs-dep1} are now given as sums.
The yield in the bottom-up thermalization scenario has the uncertainty associated with the constant $c_T$ and it is expressed by color bands. The photon yield in the hydro scenario is naturally independent of $c_T$. 

The two plots on top of Fig.~\ref{fig:comp_early} correspond to the normalization of $Q_s$ shown in Fig.~\ref{fig:Qs}, which is also indicated in Fig.~\ref{fig:Qs-dep1} by black arrows. For this choice of $Q_s$ values, at both of the RHIC and LHC energies, the yields in the two scenarios have nearly the same $N_\text{part}$ dependence. 
This agreement is accidental. For larger values of the reference $Q_s$, the bottom-up thermalization provides more photons than the hydro model extended to the early time. This is shown in the the bottom plots of Fig.~\ref{fig:comp_early}, where the normalization of $Q_s$ is chosen such that the value at the RHIC most central collision is 2 GeV. The corresponding value at the LHC is 3 GeV.

\section{Summary and outlook}
In this work, we estimated the yields for photon production from both non-equilibrium Glasma stages and the equilibrium thermal QGP stage within the bottom-up thermalization scenario of heavy-ion collisions. While the uncertainties from our lack of knowledge of the coefficients multiplying parametric estimates are large, they can be constrained significantly by the measured charged hadron multiplicity. For an assumed $Q_s^2$ profile as a function of centrality (Fig.~\ref{fig:Qs}), we found that at RHIC energies the Glasma photon yields are larger than the thermal photon yields for a wide parameter range. This dominance is especially pronounced for more peripheral collisions though we must caution that weak coupling estimates for RHIC energies are likely not reliable for peripheral collisions. At the LHC, the thermal photon yields are larger, but even at the most central collisions the Glasma contribution can range from 40--60\% of the thermal QGP yield. If the reference $Q_s$ in the $Q_s^2$ profile is increased, we find that the Glasma contribution become larger than the QGP photon yield even for central collisions. 

We also made a comparison between the photon yields in the bottom-up thermalization scenario to those in a hydro scenario that assumes the system thermalizes at the early time $\simeq Q_s^{-1}$. For the $Q_s^2$ profile shown in Fig.~\ref{fig:Qs}, the two scenarios give comparable photon yields. If we assume a larger value (by about 50\%) of the reference $Q_s$, the bottom-up thermalization scenario provides a  larger photon yield relative to the early-hydro scenario. 

Our results point to the urgent need for more refined computations to reduce the uncertainties we identified in the computation of photon yields from different stages of the Glasma evolution in addition to more sophisticated computations of thermal photon yields. These include first principles classical-statistical computations of the photon yields in the first stage of Glasma evolution as well as more detailed kinetic theory computations that match to these and to viscous hydrodynamics at later times. 

We have only considered photon yields in this work. Computations of photon spectra and flow coefficients within the bottom-up framework, and their comparison to the available data, can help either rule out the bottom-up framework of QGP equilibration or at least strongly constrain the viable set of free parameters. The possibility that these can be extracted with increasing precision offers the promise that a quantitative theory of the equilibration of strongly correlated quark and gluon matter can be developed further and tested.

\section*{Acknowledgements}
We would like to thank Oscar Garcia-Montero, Niklas Mueller, Chun Shen, Bjoern Schenke, Soeren Schlichting and Prithwish Tribedy for very valuable discussions and comments. R.~V.~is supported under DOE Contract No.~DE-{SC001}2704. 
He would like to thank the Institut f\"{u}r Theoretische Physik, Universit\"{a}t Heidelberg for kind hospitality and support via the Excellence Initiative during the early stages of this work.
This work is part of and supported by the DFG Collaborative Research Centre ``SFB~1225~(ISOQUANT)".

\appendix
\section{Small-angle approximation for the photon production} \label{sec:small-angle}
The scattering amplitude due to the exchange of a massless particle has an infrared divergence when the exchanged momentum goes to zero. 
Indeed, the amplitude for the pair annihilation process in Eq.~\eqref{eq:anni} and that of Compton scattering in Eq.~\eqref{eq:comp} diverge for $t\to 0$ or $u\to 0$. The small-angle approximation is applicable to such collision processes \cite{landau1936kinetic,lifshitz1981physical}. 
When an incoming particle and an outgoing particle that is kicked by small momentum exchange are of the same species, the collision integral for that process can be approximated as a diffusion term.
When the incoming and the outgoing particle are different species, the collision integral is replaced by a source term, which has a simple form \cite{Blaizot:2014jna}. Photon production by the pair annihilation and Compton scattering corresponds to this latter case. 

Following the same procedure as the calculation outlined in \cite{Blaizot:2014jna}, we apply the small-angle approximation to the photon production formula in Eq.~\eqref{eq:rate0}. 
First, let us consider the pair annihilation process. 
Since the $t$-channel and the $u$-channel give the same contribution, it is sufficient to compute the $t$-channel contribution alone, multiplied by a factor of two. This gives
\begin{equation} 
E\frac{dN^\text{anni}}{d^4 X d^3p} 
= \frac{1}{2(2\pi)^3} \frac{320}{9} 16\pi^2 \alpha \alpha_s \int_{p_1, p_2,p_3} \frac{u}{t} (2\pi)^4 \delta^4 (P_1 +P_2 -P_3 -P ) 
f_q (\bp_1 )f_q (\bp_2 ) \left[ 1+ f_g (\bp_3 ) \right] \, .
\end{equation}
We expand kinematic variables in terms of the exchanged momentum $\bq=\bp -\bp_1$ (henceforth asssumed to be small in magnitude) to obtain, 
\begin{gather}
p = \sqrt{(\bp_1 +\bq )^2} = p_1+\bq \cdot \bv_1 +\mathcal{O} (q^2) \, , \\
p_3 = \sqrt{(\bp_2 -\bq )^2} = p_2-\bq \cdot \bv_2 +\mathcal{O} (q^2) \, ,
\end{gather}
with $\bv_{1,2} =\bp_{1,2} /p_{1,2}$. 
Furthermore,
\begin{gather}
s = (P_1+P_2)^2 =2p_1 p_2 \left( 1-\bv_1 \cdot \bv_2 \right) \, , \\
t  = -Q^2 = -q^2 +(\bq \cdot \bv_1 )^2 +\mathcal{O} (q^3) \, , \\
u= -s-t = -s +\mathcal{O} (q^2) \, , \\
p_1 +p_2 -p_3 -p = \bq \cdot (\bv_2 -\bv_1 ) +\mathcal{O} (q^2) \, . 
\end{gather}
Keeping the leading order terms in $q$, one obtains
\begin{align}
E\frac{dN^\text{anni}}{d^4 X d^3p} 
&= \frac{20}{9\pi^3} \alpha \alpha_s \int \! d^3 q \int \! \frac{d^3 p_2}{(2\pi)^3} \frac{1}{p_2} \frac{1-\bv \cdot \bv_2}{q^2 -(\bq \cdot \bv )^2} \delta \left( \bq \cdot (\bv_2 -\bv ) \right) 
f_q (\bp )f_q (\bp_2 ) \left[ 1+ f_g (\bp_2 ) \right] \, ,
\end{align}
where, similarly to $\bv_1$ and $\bv_2$, $\bv =\bp /p$.
By a straightforward computation, one can show that the $\bq$-integration is independent of $\bv$ and $\bv_2$, and express it as 
\begin{equation}
2\pi \mathcal{L} \equiv 
\int \! d^3 q \frac{1-\bv \cdot \bv_2}{q^2 -(\bq \cdot \bv )^2} \delta \left( \bq \cdot (\bv_2 -\bv ) \right) 
= 2\pi \int \! \frac{dq}{q} \, .
\end{equation}
where the logarithmic divergence can be further expressed as 
\begin{equation}
\mathcal{L} = \int_{\Lambda_\text{IR}}^{\Lambda_\text{UV}} \frac{dq}{q} = \log \frac{\Lambda_\text{UV}}{\Lambda_\text{IR}} \, .
\end{equation}
In thermal field theory, the IR cutoff is given by the Debye mass scale $m_D\sim g^2 T^2$ and the ultraviolet cutoff is given by the temperature $T$. Hence $\mathcal{L}\sim \log(1/g)$. 

The production rate from the annihilation process simplifies in the small-angle approximation to 
\begin{equation}
E\frac{dN^\text{anni}}{d^4 X d^3p} 
= \frac{40}{9\pi^2} \alpha \alpha_s\, \mathcal{L} \, f_q (\bp ) \int \! \frac{d^3 p^\prime}{(2\pi)^3} \frac{1}{p^\prime}  
f_q (\bp^\prime ) \left[ 1+ f_g (\bp^\prime ) \right] \, .
\end{equation}
The expression within the integrand  corresponds to the density of scatterers (quarks/anti-quarks) enhanced by the Bose factor of the final state gluons. 

For the Compton scattering contribution, we can neglect the $s$-channel contribution in this approximation. 
Following the same procedure as in that for the annihilation process, one can derive,
\begin{equation}
E\frac{dN^\text{Comp}}{d^4 X d^3p} 
= \frac{40}{9\pi^2} \alpha \alpha_s \mathcal{L} \, f_q (\bp ) \int \! \frac{d^3 p^\prime}{(2\pi)^3} \frac{1}{p^\prime}  
f_g (\bp^\prime ) \left[ 1- f_q (\bp^\prime ) \right] \, .
\end{equation}
The expression within the integrand in this case corresponds to the density of scatterers (gluons) suppressed by the Pauli factor of the final state quarks/anti-quarks. 
Summing the Compton and annihilation contributions, we obtain 
\begin{equation} \label{eq:saa_total}
E\frac{dN}{d^4 X d^3p} 
= \frac{40}{9\pi^2} \alpha \alpha_s \mathcal{L} \, f_q (\bp ) \int \! \frac{d^3 p^\prime}{(2\pi)^3} \frac{1}{p^\prime}  
\left[ f_g (\bp^\prime ) +f_q (\bp^\prime ) \right] \, .
\end{equation}

Since the small-angle approximation  computation of the photon rate only involves kinematic approximations based on the dominant contributions to on-shell $2\leftrightarrow 2$ quark-gluon scattering, our result is applicable to either equilibrium or non-equilibrium situations where kinetic theory is applicable. To check the validity of this approximation, let us consider thermal equilibrium and compare our approximate result in Eq.~\eqref{eq:saa_total} with the thermal photon rate we quoted in Eq.~\eqref{eq:thermal-rate}. 
For the equilibrium distribution (with vanishing chemical potential),
\begin{equation}
\int \! \frac{d^3 p^\prime}{(2\pi)^3} \frac{1}{p^\prime}  \left[ f_g (\bp^\prime ) +f_q (\bp^\prime ) \right] 
= \frac{T^2}{8} \, .
\end{equation} 
The thermal rate with the small-angle approximation then gives 
\begin{align}
E\frac{dN^\text{th}}{d^4 X d^3p} \bigg|_\text{small-angle approx.}
&= \frac{10}{9} \frac{\alpha \alpha_s}{2\pi^2} T^2 \mathcal{L} \, f_q (p ) \notag \\
&\hspace{-5pt} \underset{p\gg T}{=} \frac{10}{9} \frac{\alpha \alpha_s}{2\pi^2} T^2 \mathcal{L} \, e^{-p/T} \, .
\end{align}
If we identify the Coulomb logarithm $\mathcal{L}$ with the logarithmic factor in Eq.~\eqref{eq:thermal-rate},  two results agree up to  a numerical factor of 2. 
We note, however, that this discrepancy can be traded for the uncertainty in ${\mathcal L}$. 


\end{document}